\newif\iffullversion
\fullversiontrue
\iffullversion
\documentclass[runningheads]{llncs}
\else
\documentclass{llncs}
\fi

\usepackage[backend=biber,giveninits=true,
style=numeric-short,doi=false,maxbibnames=100]{biblatex}
\usepackage[utf8]{inputenc}
\PassOptionsToPackage{hyphens}{url}\usepackage{hyperref}
\usepackage{microtype}
\usepackage{amssymb,latexsym}
\usepackage{amsmath}
\usepackage{mathtools}
\usepackage{xcolor}
\usepackage{booktabs}
\usepackage{float} % Allows placing floats in minipage

\usepackage{tikz}
\usetikzlibrary{myautomata}
\usetikzlibrary{extshapes}

\usepackage{pgfplots}
\definecolor{darkgreen}{rgb}{0,0.5,0}

\newcommand{\AP}{\mathit{A\hskip-0.1ex P}}
\newcommand{\suf}[1]{_{#1..}}
\newcommand{\X}{\mathsf{X}}
\DeclareMathOperator{\U}{\mathbin{\mathsf{U}}}
\DeclareMathOperator{\R}{\mathbin{\mathsf{R}}}
\newcommand{\G}{\mathsf{G}}
\newcommand{\F}{\mathsf{F}}
\newcommand{\Mod}{\mathit{Mod}}
\newcommand{\FG}{\F,\G}

\newcommand{\true}{\mathrm{{\mathit{tt}}}}
\newcommand{\false}{\mathrm{{\mathit{ff}}}}

\newcommand{\Fin}{\mathsf{Fin}}
\newcommand{\Inf}{\mathsf{Inf}}
\newcommand{\mA}{\mathcal{A}}

\newcommand{\mF}{\mathcal{F}}

\newcommand{\mM}{\mathcal{M}}

\newcommand{\mO}{\mathcal{O}}
\newcommand{\mP}{\mathcal{P}}

\newcommand{\mU}{\mathcal{U}}

\newcommand{\lang}[1]{\mathcal{L}(#1)}

\newcommand{\dm}{\mathsf{me}}

\newcommand{\oDelta}{\Delta}

\DeclareMathOperator*{\range}{{range}}
\DeclareMathOperator*{\dom}{{dom}}

\newcommand{\gm}{\tacc{3}{\phantom{0}}}
\newcommand{\om}{\tacc{2}{\phantom{0}}}

\newcommand{\bsq}{\taccsq{0}{\phantom{0}}}

\title{LTL to Smaller Self-Loop\\ Alternating Automata and
  Back\thanks{%
\iffullversion
    This is a full version of the paper accepted to ICTAC 2019.
    F.~Blahoudek has been supported by the F.R.S.-FNRS
    grant F.4520.18 (ManySynth).
\else
    F.~Blahoudek has been supported by the F.R.S.-FNRS 
    grant F.4520.18 (ManySynth). J.~Major and J.~Strej\v{c}ek
    have been supported by the Czech Science Foundation grant
    GA19-24397S.
\fi
  }}
\titlerunning{LTL to Smaller Self-Loop Alternating Automata and Back}
\author{František Blahoudek\inst{1} \and Juraj Major\inst{2} \and Jan
Strejček\inst{2}}%
\authorrunning{Blahoudek, Major, and Strejček}%
\institute{University of Mons, Belgium\and Masaryk University, Brno,
Czech Republic\\ \email{\{xblahoud,\,major,\,strejcek\}@fi.muni.cz} }

\begin{document}

\maketitle

%{{{ abstract

\begin{abstract}
  Self-loop alternating automata (SLAA) with Büchi or co-Büchi
  acceptance are popular intermediate formalisms in translations of
  LTL to deterministic or nondeterministic automata. This paper
  considers SLAA with generic transition-based Emerson-Lei acceptance
  and presents translations of LTL to these automata and
  back. Importantly, the translation of LTL to SLAA with generic
  acceptance produces considerably smaller automata than previous
  translations of LTL to Büchi or co-Büchi SLAA. Our translation is
  already implemented in the tool LTL3TELA,
  where it helps to produce
  small deterministic or nondeterministic % transition-based Emerson-Lei
  automata for given LTL formulae.
\end{abstract}

%}}}

%{{{ intro

\section{Introduction}

Translation of \emph{linear temporal logic
  (LTL)}~\cite{pnueli.77.focs} into equivalent automata over infinite
words is an important part of many methods for model checking,
controller synthesis, monitoring, etc. %, vacuity checking etc.
This paper presents improved translations of LTL to \emph{self-loop
  alternating automata (SLAA)}~\cite{tauriainen.06.phd}, which are
alternating automata that contain no cycles except self-loops. The SLAA class
% SLAA with Büchi or co-Büchi acceptance
is studied for more than 20 years under several different names
including \emph{very weak}~\cite{rohde.97.phd,gastin.01.cav},
\emph{linear}~\cite{loeding.00.ifip}, \emph{linear
  weak}~\cite{hammer.05.tacas}, or
\emph{1-weak}~\cite{pelanek.05.ciaa} \emph{alternating automata}. The
first publications showing that any LTL formula can be easily
translated to an SLAA with only a linear number of states in the
length of the formula are even
older~\cite{muller.88.lics,vardi.94.tacs}. An LTL to SLAA translation
forms the first step of many LTL to automata translations. For
example, it is used in popular tools LTL2BA~\cite{gastin.01.cav} and
LTL3BA~\cite{babiak.12.tacas} translating LTL to nondeterministic
automata, and also in the tool LTL3DRA~\cite{babiak.13.atva}
translating a fragment of LTL to deterministic automata.

A nice survey of various instances of LTL to SLAA translations can be
found in Tauriainen's doctoral thesis~\cite{tauriainen.06.phd}, where
he also presents another improved LTL to SLAA translation. To our best
knowledge, the only new improvement since publication of the thesis
has been presented by Babiak et al.~\cite{babiak.12.tacas}. All the
LTL to SLAA translations considered so far produce SLAA with
(state-based) Büchi or co-Büchi acceptance. The only exception is
the translation by Tauriainen producing SLAA with transition-based
co-Büchi acceptance.

In this paper, we follow a general trend of recent research and
development in the field of automata over infinite words and their
applications: consider a more general acceptance condition to
construct smaller automata. In theory, this change usually does not
decrease the upper bound on the size of constructed
automata. Moreover, the complexity of algorithms processing automata
with a more involved acceptance condition can be even higher. However,
practical experiences show that achieved reduction of automata size
often outweighs complications with a more general acceptance
condition. This can be documented by observations of nondeterministic
as well as deterministic automata.

Nondeterministic automata are traditionally considered with Büchi
acceptance. However, all three most popular LTL to nondeterministic
automata translators, namely LTL2BA~\cite{gastin.01.cav},
LTL3BA~\cite{babiak.12.tacas}, and Spot~\cite{duret.16.atva},
translate LTL formulae to \emph{transition-based generalized Büchi
  automata (TGBA)}, which are further transformed to Büchi
automata.  When solving emptiness check, which is the central part of
many model checking tools, algorithms designed for TGBA perform better
than algorithms analyzing the corresponding Büchi
automata~\cite{couvreur.05.spin,renault.15.tacas}.

% Situation around deterministic automata is more dramatic.
Deterministic automata were typically considered with Rabin or Streett
acceptance. Tools of the Rabinizer family~\cite{kretinsky.18.cav} and
the tool LTL3DRA~\cite{babiak.13.atva} produce also deterministic
automata with transition-based generalized Rabin acceptance. The
equivalent Rabin automata are often dramatically larger. % , which are
% frequently smaller than Rabin automata
% obtained by the traditional way, i.e.~by translation of LTL formulae
% to nondeterministic automata and subsequent determinization
% implemented for example in ltl2dstar~\cite{klein.07.ciaa}.
Direct processing of generalized Rabin automata can be substantially
more efficient as shown by Chatterjee et al.~\cite{chatterjee.13.cav}
for probabilistic model checking and LTL synthesis.

All the previously mentioned acceptance conditions 
% including the generalized ones 
can be expressed by a generic acceptance condition originally
introduced by Emerson and Lei~\cite{emerson.87.scp} and recently
reinvented in the \emph{Hanoi omega-automata (HOA)
  format}~\cite{babiak.15.cav}. Emerson-Lei acceptance condition is
any positive boolean formula over terms of the form $\Inf\gm$ and
$\Fin\gm$, where $\gm$ is an \emph{acceptance mark}. A run of a
nondeterministic automaton (or an infinite branch of a run of an
alternating automaton) satisfies $\Inf\gm$ or $\Fin\gm$ if it visits
the acceptance mark $\gm$ infinitely often or finitely often,
respectively.
% Hence, $\Inf\gm$ corresponds to Büchi acceptance while $\Fin\gm$ corresponds to co-Büchi acceptance.
The acceptance marks placed on states denote traditional state-based
acceptance, while marks placed on transitions correspond to
transition-based acceptance.

Some tools that work with \emph{transition-based Emerson-Lei automata
(TELA)} already exist. For example, Delag~\cite{muller.17.gandalf}
produces deterministic TELA and Spot is now able to produce both
deterministic and nondeterministic TELA. The produced TELA are often
smaller than the automata produced by the tools mentioned in the
previous paragraphs.
%
%For nondeterministic automata, this can be explained by the generic
%acceptance as TELA can be obviously more succinct than TGBA. This is
%not the case for deterministic automata, as every deterministic TELA
%can be transformed into an equivalent deterministic transition-based
%generalized Rabin automaton with the same number of states and edges
%(only the acceptance condition can be exponentially bigger). 
%
The development version of Spot provides also an emptiness check for
TELA, and a probabilistic model checking algorithm working with
deterministic Emerson-Lei automata has been implemented in PRISM. In
both cases, an improved performance over previous solutions has been
reported~\cite{baier.19.atva}.

This paper presents a translation of LTL to SLAA with transition-based
Emerson-Lei acceptance. The translation aims to take advantage of the
generic acceptance and produce SLAA with less states. We
present it in three steps.
% actually three translations of LTL to SLAA with
% increasing complexity of acceptance condition.
\begin{enumerate}
\item Section~\ref{sec:basic} recalls a basic translation producing
  co-Büchi SLAA. The description uses the same terminology and
  notation as the following modified translations. In particular, the
  acceptance marks are on transitions.
\item In Section~\ref{sec:f-merging}, we modify the translation such
  that states for subformulae of the form $\F\psi$ are merged with
  states for~$\psi$. The technique is called \emph{$\F$-merging}. The
  acceptance condition of constructed SLAA is a positive
  boolean combination of $\Fin$-terms. We call such automata
  \emph{$\Inf$-less SLAA}.
\item Finally, we further modify the translation in
  Section~\ref{sec:fg-merging}, where states for some subformulae of
  the form $\G\psi$ are merged with states for~$\psi$. The resulting
  technique is thus called \emph{$\FG$-merging}. Constructed SLAA 
  use acceptance condition containing both $\Inf$- and $\Fin$-terms.
% in a specific structure called \emph{sibling acceptance}.
\end{enumerate}
The difference between these translations is illustrated by
Figure~\ref{fig:SLAA-demo} showing three SLAA for the formula
$\F(\G a \vee \G\F b)$. One can observe that the initial state of the
middle automaton is merged with the states for $\G a$ and $\G\F b$ due
to $\F$-merging. In the automaton on the right, the state for $\G\F b$
is merged with $\F b$ and the initial state is then merged with $\G a$
and $\G\F b$. Hence, the resulting automaton contains only one state
and the LTL to SLAA translation in this case produces directly a
nondeterministic automaton.

\begin{figure}[t]
\centering
\begin{tikzpicture}[automaton]

\begin{scope}[xshift=-4.5cm,initial angle=90]
\node[state,initial] (phi) at (0,0) {$\varphi\vphantom{\G}$};
\node[cstate] (Ga) at (1.5,0) {$\G a$};
\node[cstate] (GFb) at (0,-1.5) {$\G\F b$};
\node[state] (Fb) at (1.5,-1.5) {$\F b$};
\path[->,auto]
(phi) edge[loop left]
  pic {acc={3}{\phantom{0}}} pic {l=$\true$} (phi)
(phi) edge node[above]{$a$} (Ga)
(Ga) edge[loop right] node{$a$} (Ga)
(phi) edge node[left]{$b$} (GFb)
(phi) edge[out=-45,in=45] (GFb)
(phi) edge[out=-45,in=105]
  node[above right,inner sep=.3,pos=.45]{$\bar{b}$} (Fb)
(GFb) edge[loop left] node{$b$} (GFb)
(GFb) edge node[]{$\bar{b}$} (Fb)
(GFb) edge[out=0,in=-30,looseness=7] (GFb)
(Fb) edge[loop right]
  pic {acc={3}{\phantom{0}}} pic {l=$\bar{b}$} (phi)
(Fb) edge node {$b$} + (0,-1.1)
;
\node[acclabel] at (.75,-3.2) {$\Fin\gm$};
\node at (.75,-4) {(basic)};
\end{scope}

\begin{scope}[initial angle=90]
\node[state,initial] (phi) at (0,0) {$\varphi\vphantom{\G}$};
\node[state] (Fb) at (1.5,-1.5) {$\F b$};
\path[->,auto]
(phi) edge[loop left]
  pic {acc={3}{1}} pic {l=$\true$} (phi)
(phi) edge[loop right]
  pic {acc={1}{3}} pic {l=$a$} (phi)
(phi) edge[out=-135,in=-105,looseness=10]
  pic[pos=.12] {acc={2}{2}} pic[pos=.3,left] {l=$b$} (phi)
%(phi) edge[out=-45,in=45] (Fb)
(phi) edge[out=-45,in=105]
  node[above right,inner sep=.3,pos=.45] {$\bar{b}$} (Fb)
(phi) edge[out=-45,in=-75,looseness=10]
  pic[pos=.12] {acc={2}{2}} (phi)
(Fb) edge[loop right]
  pic {acc={3}{1}} pic {l=$\bar{b}$} (phi)
(Fb) edge node {$b$} + (0,-1.1)
;
\node[acclabel] at (.75,-3.2)
  {$\Fin\tacc{3}{1}\wedge(\Fin\tacc{2}{2}\vee\Fin\tacc{1}{3})$};
\node at (.75,-4) {($\F$-merging)};
\end{scope}

\begin{scope}[xshift=4.8cm,initial angle=90]
\node[state,initial] (phi) at (0,-.7) {$\varphi\vphantom{\G}$};
\path[->,auto]
(phi) edge[loop left] 
  pic {acc={3}{1}} pic {l=$\true$} (phi)
(phi) edge[loop right]
  pic {acc={1}{3}} pic {l=$a$} (phi)
(phi) edge[out=-135,in=-105,looseness=10]
  pic[pos=.75] {accsq={0}{5}} pic[pos=.15] {acc={2}{2}}
  pic[left] {l=$b$} (phi)
(phi) edge[out=-45,in=-75,looseness=10]
  pic[pos=.15] {acc={2}{2}} pic[pos=.75] {acc={4}{4}}
  pic[right] {l=$\bar{b}$} (phi)
;
\node[acclabel,align=center] at (0,-2.8)
  {$\Fin\tacc{3}{1}\wedge{}$\\$(\Fin\tacc{2}{2}\vee\Fin\tacc{1}{3})
   \wedge{}$\\$(\Fin\tacc{4}{4}\vee\Inf\taccsq{0}{5})$};
\node at (0,-4) {($\FG$-merging)};
\end{scope}

% \begin{scope}[xshift=4.7cm]
% \node[cstate,initial,text width=1.2cm] (f) at (0,-0.6) {};
% \path[->,auto,looseness=9]
% (f.25) edge[out=55,in=35,] pic{acc={1}{1}} pic[pos=.19]{acc={2}{2}} pic[above]{l=$\bar{a}\bar{b}\bar{c}$} (f.10)
% (f.-170) edge[out=-145,in=-125,] pic{acc={1}{1}} pic[pos=.81]{acc={0}{3}} pic[below]{l=$\bar{a}bc$} (f.-155)
% (f.60) edge[looseness=9,out=100,in=85,outer sep=3pt] pic[pos=.19]{acc={2}{2}} pic{l=$a\bar{b}\bar{c}$} (f.40)
% (f.140) edge[looseness=9,out=100,in=85,outer sep=3pt] pic{acc={1}{1}} pic[pos=.19]{acc={2}{2}} pic[pos=.81]{acc={0}{3}} pic{l=$\bar{a}\bar{b}c$} (f.120)
% (f.170) edge[,out=145,in=125,outer sep=3pt] pic[pos=.19]{acc={2}{2}} pic[pos=.81]{acc={0}{3}} pic[above]{l=$a\bar{b}c$} (f.155)
% (f.-25) edge[,out=-55,in=-35,outer sep=3pt] pic[below]{l=$ab\bar{c}$} (f.-10)
% (f.-60) edge[,out=-100,in=-80,outer sep=4pt] pic{acc={1}{1}} pic[below]{l=$\bar{a}b\bar{c}$} (f.-40)
% (f.-140) edge[,out=-100,in=-80,outer sep=3pt] pic[pos=.81]{acc={0}{3}} pic[below]{l=$abc$} (f.-120)
% ;
% \node[acclabel] at (-.1,-2.4) {$\Fin\tacc{1}{1} \lor \Fin\tacc{2}{2}\lor\Inf\tacc{0}{3}$};
% \end{scope}

\end{tikzpicture}
\caption{Automata for the formula $\varphi=\F(\G a \vee\G\F b)$:
  the co-Büchi SLAA produced by the basic translation (left), the
  $\Inf$-less SLAA produced by $\F$-merging (middle), and the SLAA
  produced by $\FG$-merging (right). Graphical notation is explained
  in Section~\ref{sec:prelim}.}
\label{fig:SLAA-demo}
\end{figure}

LTL to SLAA translations are traditionally accompanied by automata
simplification based on transition
dominance~\cite{gastin.01.cav}. Section~\ref{sec:simplification}
extends this idea to SLAA with generic acceptance.

Section~\ref{sec:slaa2ltl} completes the theoretical part of the paper
with a backward translation which takes an SLAA with transition-based
Emerson-Lei acceptance and produces an equivalent LTL
formula. Altogether, we get that SLAA with the generic acceptance have
the same expressiveness as LTL.

The three presented LTL to SLAA translations have been implemented and
Section~\ref{sec:experiments} provides their experimental comparison
extended with the results of the LTL to SLAA translation implemented
in LTL3BA. On randomly generated formulae containing only temporal
operators $\F$ and $\G$, which are favourable to our translation
improvements, the $\FG$-merging can save over 45\% of states. This is
a considerable reduction, especially with respect to the fact that
even the simplest LTL to SLAA translations produce automata of linear
size and thus the space for reduction is not big.
 
As we said at the beginning, SLAA are mainly used as an
intermediate formalism in translations of LTL to other kinds of
automata. We have already developed a \emph{dealternation} algorithm
transforming the produced SLAA to nondeterministic TELA. Both
$\FG$-merging translation and dealternation are implemented in
LTL3TELA~2.1~\cite{major.19.atva}, which combines them with some
heuristics and many functions of Spot in order to translate LTL to
small deterministic and nondeterministic
automata. Experiments\footnote{\url{https://github.com/jurajmajor/ltl3tela/blob/master/ATVA19.md}}
show that the tool produces, on average, smaller automata than
established state-of-the-art translators.

% As a result, LTL3TELA produces, on
% average, smaller deterministic automata than other state-of-the-art
% translators including Spot, Delag, and Rabinizer~4. The situation is
% similar for nondeterministic automata and state-of-the-art translators
% Spot and LTL3BA, but the differences are smaller. These experimental
% comparisons are also presented in Section~\ref{sec:experiments}.
%
% For many formulae, this translation
% produces the smallest automaton out of all modern LTL to
% non-alternating automata translators (LTL3BA, Spot, Rabinizer 3,
% Delag). For example, for the formula $\F\G a \lor (b \land \F\neg a)$
% it creates a co-Büchi automaton with 3 states while all the other
% tools need at least 4 states. 

%}}}
%{{{ preliminaries

\section{Preliminaries}\label{sec:prelim}
This section recalls the notion of \emph{linear temporal
  logic}~\cite{pnueli.77.focs} and the definition of \emph{self-loop
  alternating automata}~\cite{tauriainen.06.phd}. We always use
automata with transition-based acceptance condition given in the
format of Emerson-Lei acceptance. For example, co-Büchi automaton is
an automaton with acceptance condition $\Fin\gm$.

\subsection{Linear Temporal Logic (LTL)}

We define the syntax of LTL formulae directly in the positive normal form as
%\[
%%  \varphi::=\true\mid\false\mid a\mid\neg a\mid\varphi\vee\varphi\mid\varphi\wedge\varphi\mid\X\varphi\mid\varphi\U\varphi\mid\varphi\R\varphi\textrm{,}
%  \varphi::=\true\,\mid\,\false\,\mid\,a\,\mid\,\neg a\,\mid\,\varphi\vee\varphi\,\mid\,\varphi\wedge\varphi\,\mid\,\X\varphi\,\mid\,\varphi\U\varphi\,\mid\,\varphi\R\varphi\textrm{,}
%\]
\[
\varphi::=\true\,\mid\,\false\,\mid\,a\,\mid\,\neg
a\,\mid\,\varphi\vee\varphi\,\mid\,\varphi\wedge\varphi\,\mid\,\X\varphi\,\mid\,\varphi\U\varphi\,\mid\,\varphi\R\varphi,
\]
where $\true$ stands for \emph{true}, $\false$ for \emph{false}, $a$
ranges over a set $\AP$ of \emph{atomic propositions}, and
$\X,\U,\R$ are temporal operators called \emph{next}, \emph{until},
and \emph{release}, respectively. 
% An \emph{alphabet} is a set
% $\Sigma=2^{\AP'}$, where $\AP'$ is a finite subset of $\AP$. A
% \emph{word} over $\Sigma$ is a % an infinite
% sequence $u=u_0u_1u_2\ldots\in\Sigma^\omega$.
A \emph{word} is an infinite
sequence $u=u_0u_1u_2\ldots\in\Sigma^\omega$, where $\Sigma=2^\AP$. By
$u\suf{i}$ we denote the suffix $u\suf{i}=u_iu_{i+1}\ldots$.  We
define when a word $u$ \emph{satisfies} $\varphi$, written
$u\models\varphi$, as follows:
\begin{tabbing}
  \hspace*{1em} \= $u\models\true$\\
  \> $u\models a$ \hspace*{2.7em} \= iff~ \= $a\in u_0$\\
  \> $u\models \neg a$ \> iff \> $a\not\in u_0$\\
  \> $u\models\varphi_1\vee\varphi_2$ \> iff \> $u\models\varphi_1$ or $u\models\varphi_2$\\
  \> $u\models\varphi_1\wedge\varphi_2$ \> iff \> $u\models\varphi_1$ and $u\models\varphi_2$\\
  \> $u\models\X\varphi$ \> iff \> $u\suf{1}\models\varphi$\\
  \> $u\models\varphi_1\U\varphi_2$ \> iff \>
     $\exists i\ge 0$ such that $u\suf{i}\models\varphi_2$ and
     $\forall\, 0\leq j<i\,.~u\suf{j}\models\varphi_1$\\
  \> $u\models\varphi_1\R\varphi_2$ \> iff \>
     $\exists i\ge 0$ such that $u\suf{i}\models\varphi_1$ and
     $\forall\, 0\le j\le i\,.~u\suf{j}\models\varphi_2$,\\
  \> \> \> or $\forall i\ge 0\,.~u\suf{i}\models\varphi_2$
\end{tabbing}
%\end{center}

A formula $\varphi$ defines the language
$L(\varphi)=\{u\in(2^{\AP(\varphi)})^\omega\mid u\models\varphi\}$, where
$\AP(\varphi)$ denotes the set of atomic propositions occurring in $\varphi$.
Further, we use derived operators \emph{eventually} ($\F$) and
\emph{always} ($\G$) defined by $\F\varphi\equiv\true\U\varphi$ and
$\G\varphi\equiv\false\R\varphi$. A \emph{temporal
  formula} is a formula where the topmost operator is neither
conjunction, nor disjunction. A formula without any temporal operator
is called \emph{state formula}. Formulae $\true,\false,a,\neg a$ are
both temporal and state formulae.

\subsection{Self-Loop Alternating Automata (SLAA)} 

An \emph{alternating automaton} is a tuple
$\mA = (S,\Sigma,\mM,\Delta,s_I,\varPhi)$, where 
\begin{itemize}
\item $S$ is a finite set of \emph{states}, 
\item $\Sigma$ is a finite \emph{alphabet},
\item $\mM$ is a finite set of \emph{acceptance marks},
\item $\Delta \subseteq S \times \Sigma \times 2^{\mM} \times 2^S$ is
  an \emph{alternating transition relation},
\item $s_I\in S$ is the \emph{initial state}, and 
\item $\varPhi$ is an \emph{acceptance formula}, which is a positive
  boolean combination of terms $\Fin\gm$ and $\Inf\gm$, where $\gm$
  ranges over $\mM$.
\end{itemize}
An alternating automaton is a \emph{self-loop alternating automaton
  (SLAA)} if there exists a partial order on $S$ such that for every
$(s,\alpha,M,C) \in \Delta$, all states in $C$ are lower or equal to
$s$. In other words, SLAA contain no cycles except self-loops.

Subsets $C\subseteq S$ are called \emph{configurations}.  A
quadruple $t = (s,\alpha,M,C)\in\Delta$ is called a \emph{transition}
from $s$ to $C$ under $\alpha$ (or labelled by $\alpha$ or
$\alpha$-transition) marked by elements of $M$. A transition
$t=(s,\alpha,M,C)\in\Delta$ is \emph{looping} (or simply a
\emph{loop}) if its \emph{destination configuration} $C$ contains its
\emph{source}~$s$.

A \emph{multitransition} $T$ under $\alpha$ is a set of transitions under
$\alpha$ such that the source states of the transitions are pairwise different.
%The \emph{source configuration} of $T$, denoted by $\dom(T)$, is the
%set of source states of transitions in $T$.  The \emph{destination
% configuration}
%of $T$, denoted by $\range(T)$, is the union of destination configurations of
%transitions in $T$. The \emph{acceptance label} of $T$, denoted as $\acc(T)$,
% is
%the union of acceptance labels of the transitions in $T$.
%
The \emph{source configuration} $\dom(T)$ of a multitransition $T$ is
the set of source states of transitions in $T$. The
\emph{destination configuration} $\range(T)$
% and the \emph{acceptance marks} $M(T)$ of $T$
is the union of destination configurations
% and acceptance marks
of transitions in $T$.
% , respectively.
For an alternating automaton $\mA$, $\Gamma^\mA$ denotes the set of
all multitransitions of $\mA$ and $\Gamma_\alpha^\mA$ denotes the set
of all multitransitions of $\mA$ under~$\alpha$.
% We write just $\Gamma$ and $\Gamma_\alpha$ when
% $\mA$ is clear from the context.

A \emph{run} $\rho$ of an alternating automaton $\mA$ over a word
$u = u_0u_1\ldots \in \Sigma^\omega$ is an infinite sequence
$\rho=T_0T_1\ldots$ of multitransitions such that $\dom(T_0)=\{s_I\}$
and, for all $i\geq0$, $T_i$ is labelled by $u_i$ and
$\range(T_i)=\dom(T_{i+1})$. Each run $\rho$ defines a directed
acyclic edge-labelled graph $G_\rho = (V, E, \lambda)$, where
%$V=\bigcup_{i=0}^{\infty} V_i$ with $V_i = \dom(T_i)\times\{i\}$,
%\begin{itemize}
%\item $V = \bigcup_{i=0}^{\infty} V_i$, where $V_i = \dom(T_i)\times\{i\}$,
%\item $E = \bigcup_{i=0}^\infty \left\{\big((s,i),(s',i+1)\big) \mid
%                                (s,\alpha,A,C) \in T_i, s'\in C \right\}$, and
%\item $\lambda\big((v_1,v_2)\big) = A$ if $v_1 = (s,i), v_2 = (s',i+1)$
%                             for some $i$ and
%                             $(s,\alpha,A,C) \in T_i$ 
%                             and $s'\in C$, 
%                             and is undefined otherwise\footnote{
%                             Note that there is at most one transition
%                             of $s$ in $T_i$ by definition of multitransition and
%                             thus $\lambda$ is indeed a function.},\myquestion{label only edges $((s,i),(s,i+1))$?}
%\end{itemize}
%
\begin{align*}
V &= \bigcup_{i=0}^{\infty} V_i\text{, where }V_i = \dom(T_i)\times\{i\},\\
E &= \bigcup_{i=0}^\infty \left\{\big((s,i),(s',i+1)\big) \mid
                                (s,\alpha,M,C) \in T_i, s'\in C \right\}\text{, and}
\end{align*}
the labeling function $\lambda:E\rightarrow2^{\mM}$ assigns to each
edge $e=((s,i),(s',i+1)\big)\in E$ the acceptance marks from the
corresponding transition, i.e., $\lambda(e)=M$ where
$(s,\alpha,M,C)\in T_i$. %
A \emph{branch} of the run $\rho$ is a maximal (finite or infinite)
sequence $b=(v_0,v_1)(v_1,v_2)\ldots$ of consecutive edges in $G_\rho$
such that $v_0 \in V_0$. For an infinite branch $b$, let $M(b)$ denote
the set of marks that appear in infinitely many sets of the sequence
$\lambda((v_0,v_1))\lambda((v_1,v_2))\ldots$. An infinite branch $b$
satisfies $\Inf\gm$ if $\gm\in M(b)$ and it satisfies $\Fin\gm$ if
$\gm\not\in M(b)$. An infinite branch is \emph{accepting} if it
satisfies the acceptance formula $\varPhi$. We say that a run $\rho$
is \emph{accepting} if all its infinite branches are accepting. The
language of $\mA$ is the set $L(\mA)=\{u\in\Sigma^\omega\mid\mA\text{
has an accepting run over }u\}$.

\medskip Several examples of SLAA are given in
Figure~\ref{fig:SLAA-demo}. Examples of SLAA with their runs can be
found in Figure~\ref{fig:fg-merging:motivation}. Note that a
transition $(s,\alpha,M,C)\in\Delta$ of an alternating automaton is
visualised as a branching edge leading from $s$ to all states in
$C$. In this paper, an automaton alphabet has always the form
$\Sigma=2^{\AP'}$, where $AP'$ is a finite set of atomic
propositions. To keep the visual representation of automata concise,
edges are labelled with boolean formulae over atomic propositions in a
condensed notation: $\bar{a}$ denotes $\neg a$ and conjunctions are
omitted. Hence, $ab\bar{c}$ would stand for $a\land b\land\neg
c$. Every edge represents all transitions under combinations of atomic
propositions satisfying its label.

% We can visualize run DAGs as in Figure~\ref{fig:vwaarun} which represents a run
% of the VWAA $\mA$ of Figure~\ref{fig:vwaa}. The dotted lines through nodes of
% levels divide the DAG into segments corresponding to multitransitions.  All
% nodes in one row are labelled by the same state which is depicted at their
% left-hand side. Each transition $t$ of a multitransition is represented by edges
% leading across the corresponding segment from the node labelled by the source
% state of $t$ to nodes labelled by states of the destination configuration of
% $t$. As our alternating automata are very weak, we can order the states in a way
% that all edges in any DAG go only to the same or a lower row.

%For each index $j\in J$ we can classify runs into three categories. The run $\rho$ is 
%\begin{description}
%\item[$j$-fin] if $j\in A_i$ for finite number of $i$, or
%\item[$j$-inf-$\branches$-fin]\myquestion{$j$-inf-branch-fin?} if there are infinitely
%many $i$ such that $j\in A_i$ but each branch of $G_\rho$ contains only finite
%number of edges with $j$ in label, or
%\item[$j$-inf-$\branches$-inf] if there are infinitely many $i$ such that $j\in A_i$
%and some branch of $G_\rho$ contains infinitely many edges with $j$ in label.
%\end{description}

%}}}
%{{{ basic translation

\section{Basic Translation}\label{sec:basic}

This section presents a basic translation of LTL to co-Büchi SLAA
similar to the one implemented in LTL3BA~\cite{babiak.12.tacas}. To
simplify the presentation, in contrast to the translation of LTL3BA we
omit the optimization called \emph{suspension}, we describe
transitions for each $\alpha\in\Sigma$ separately, and we slightly
modify the acceptance condition of the SLAA; in particular, we switch
from state-based to transition-based acceptance.
% The proof of Theorem~\ref{thm:basic} by structural induction can be found in Major's thesis~\cite{major.17.thesis}.

Let $\varphi$ be an LTL formula, where subformulae of the form
$\F\psi$ and $\G\psi$ are seen only as abbreviations for
$\true\U\varphi$ and $\false\R\varphi$, respectively. An equivalent
SLAA is constructed as
$\mA_\varphi = (S, \Sigma, \{\gm\}, \Delta, \varphi, \Fin\gm)$, where
states in $S$ are subformulae of $\varphi$ and
$\Sigma = 2^{\AP(\varphi)}$. The construction of the transition
relation $\Delta$ treats it equivalently as a function
$\Delta: S\times\Sigma \rightarrow 2^\mP$ where
$\mP = 2^{\mM} \times 2^{S}$.  The construction of $\Delta$ is defined
inductively and it directly corresponds to the semantics of LTL. The
acceptance mark $\gm$ is used to ensure that an accepting run cannot
stay in a state $\psi_1\U\psi_2$ forever. In other words, it ensures
that $\psi_2$ will eventually hold. The translation uses an auxiliary
product operator $\otimes$ and a marks eraser $\dm$ defined for each
$P,P'\subseteq \mP$ as:
\begin{align*}
  P\otimes P' &= \{(M \cup M', C \cup C') \mid (M,C)\in P,(M',C')\in P'\}\\
  \dm(P) &= \{(\emptyset,C) \mid (M,C) \in P\}
\end{align*}
The product operator is typically used to handle conjunction: to get
successors of $\psi_1\wedge\psi_2$, we compute the successors of
$\psi_1$ and the successors of $\psi_2$ and combine them using the
product operator $\otimes$. The marks eraser has two applications.
First, it is used to remove unwanted acceptance marks on transitions
looping on states of the form $\psi_1\R\psi_2$.
% , which would be
% inherited from transitions of $\psi_2$.
Second, it is used to remove irrelevant accepting marks from the
automaton, which are all marks not lying on loops. Indeed, only
looping transition can appear infinitely often on some branch of an
SLAA run and thus only marks on loops are relevant for acceptance.

\begin{align*}
  \Delta(\true,\alpha) &= \{(\emptyset,\emptyset)\}	\\
  \Delta(\false,\alpha) &= \emptyset \\
  \Delta(a,\alpha) &= \{(\emptyset,\emptyset)\}
       \text{ if } a\in\alpha,~ \emptyset\text{ otherwise}\\
  \Delta(\neg a,\alpha) &= \{(\emptyset,\emptyset)\}
       \text{ if } a\notin\alpha,~ \emptyset\text{ otherwise}\\
  % % % Boolean
  \Delta(\psi_1 {\land} \psi_2,\alpha) &=
       \dm\big(\Delta(\psi_1,\alpha) \otimes \Delta(\psi_2,\alpha)\big)\\
  \Delta(\psi_1 {\lor} \psi_2,\alpha) &=
       \dm\big(\Delta(\psi_1,\alpha) \cup \Delta(\psi_2,\alpha)\big)\\
  % % % Next
  \Delta(\X\psi,\alpha) &= \big\{(\emptyset,\{\psi\})\big\}\\
  % % % Until
  \Delta(\psi_1 {\U} \psi_2,\alpha) &= 
     \dm\big(\Delta(\psi_2,\alpha)\big) \cup
     \Big(\big\{(\{\!\gm\!\}, \{\psi_1 {\U} \psi_2\})\big\}
     \!\otimes\!
     \dm\big(\Delta(\psi_1,\alpha)\big)\Big)\\
  % % % Release 
  \Delta(\psi_1 {\R} \psi_2,\alpha) &= 
      \dm\big(\Delta(\psi_1,\alpha){\otimes}\Delta(\psi_2,\alpha)\big)\cup\dm\Big(\big\{(\emptyset,\{\psi_1{\R}\psi_2\})\big\}\otimes\Delta(\psi_2,\alpha)\Big)
\end{align*}

The automaton $\mA_\varphi$ has at most $|\varphi|$ states as the
states are subformulae of $\varphi$.  To prove that the constructed
automaton is a self-loop alternating automaton, it is enough to
consider the partial order \emph{`being a subformula of'} on states.

\section{$\F$-Merging Translation}
\label{sec:f-merging}

Now we modify the basic translation on subformulae of the form
$\F\psi$. The modified translation produces \emph{$\Inf$-less SLAA},
which are SLAA without $\Inf\bsq$ terms in acceptance
formulae.
%  Further, we describe a procedure that converts an $\Inf$-less SLAA with
% into a nondeterministic automaton.

% \subsection{LTL to $\Inf$-less SLAA}\label
% {sec:f-merging:SLAA}

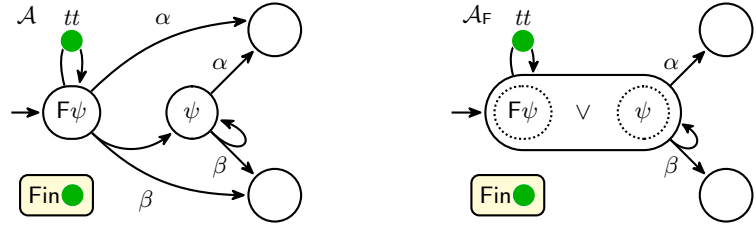
\begin{figure}[t]
\centering
\begin{tikzpicture}[automaton,xshift=6cm]
\def\psiDist{1.6}
\def\psiAx{2.7}
\def\psiBx{\psiDist}
\def\psiBy{1.1}
\def\boxPad{.5}
\def\shorten{5pt}
\def\shortenOut{4pt}
\def\nameX{-.6}
\def\nameY{1.3}
\def\labelX{-.7}
\def\labelY{-\psiBy}
% % % LEFT % % %
\begin{scope}
  \node[cstate,initial] (f) at (0,0) {$\F\psi$};
  \node[cstate] (p2) at  (\psiDist,0) {$\,\psi\,$};
  \node at (\nameX,\nameY) {$\mA$};

  \begin{scope}
    \node[state] (a2) at (\psiAx,\psiBy)  {};  
    \node[state] (b2) at (\psiAx,-\psiBy)  {};
  \end{scope}
  
  \path[->]
    (p2) edge pic[above,pos=.2]{l=$\alpha$}(a2)
    (p2) edge pic[below,pos=.2]{l=$\beta$}(b2)
    (p2) edge[out=-45,in=-15,looseness=9] (p2)
    (f)  edge[loop above,looseness=11] 
         pic{acc={3}{\phantom{0}}} pic{l=$\true$} (f)
  ;
  
  \path[->]
    (f) edge[out=-45,in=180] node[below left,pos=.5]{$\beta$} (b2)
    (f) edge[out=-45,in=-135,] (p2)
    (f) edge [bend left=20] pic[above]{l=$\alpha$} (a2.160)
  ;
  \node[acclabel,anchor=west] at (\labelX,\labelY) {$\Fin\gm$};
\end{scope}

% % % RIGHT % % %
\begin{scope}[xshift=6cm]
  \node[cstate,unreachable] (f) at (0,0) {$\F\psi$};
  \node[cstate,unreachable] (p2) at  (\psiDist,0) {$\,\psi\,$};
  \node at (.8,0) {$\lor$};
    \node at (\nameX,\nameY) {$\mA_\F$};
  \begin{scope}[]
    \node[state] (a2) at  (\psiAx,\psiBy)  {};
    \node[state] (b2) at  (\psiAx,-\psiBy)  {};
  \end{scope}
  \path[->,shorten <=\shortenOut]
  (p2) edge
       pic[above,pos=.2]{l=$\alpha$}(a2)
  (p2) edge
       pic[below,pos=.2]{l=$\beta$}(b2)
  (p2) edge[out=-45,in=-15,looseness=9,
            shorten >=\shorten] (p2)
  (f)  edge[loop above,looseness=11,
            shorten >=\shorten]
       pic{acc={3}{\phantom{0}}} pic{l=$\true$} (f)
;
  \draw[rounded corners=5mm,solid] (f) + (-\boxPad,\boxPad)
   rectangle ($(p2) + (\boxPad,-\boxPad)$);
  \draw (f) + (-.95,0) edge[->,shorten >=4.5] (f);
  \node[acclabel,anchor=west] at (\labelX,\labelY) {$\Fin\gm$};
\end{scope}
\end{tikzpicture}
\caption{Automata for $\F\psi$: the SLAA $\mA$ built by the basic translation (left) and the SLAA $\mA_\F$
built by the $\F$-merging translation (right).}
\label{fig:f-merging:motivation}
\end{figure}

Before giving the formal translation, we discuss three examples to
explain the ideas behind $\F$-merging. We start with a formula
$\F\psi$ where $\psi$ is a temporal formula. Further, assume that the
state $\psi$ of the SLAA constructed by the basic translation has two
types of transitions: non-looping labelled by $\alpha$ and loops
labelled by $\beta$. The SLAA $\mA$ for $\F\psi$ can be found in
Figure~\ref{fig:f-merging:motivation} (left).
% The state $\F\psi$ has the same transitions and
% a $\true$ labelled loop marked by \gm.
% The $\true$-loop can be taken
% only finitely many times by any accepting run of $\mA$ and therefore,
States $\F\psi$ and $\psi$ can be merged into a single state that
represents their disjunction (which is equivalent to $\F\psi$) as
shown by the SLAA $\mA_\F$ of Figure~\ref{fig:f-merging:motivation}
(right). The construction is still correct: (i) Clearly, each sequence
of transitions that can be taken in $\mA$ can be also taken in
$\mA_\F$. (ii) The sequences of transitions of $\mA_\F$ that cannot be
taken in $\mA$ are those where the $\true$-loop is taken after some
$\beta$-loop. However, every accepting run of $\mA_\F$ use the
$\true$-loop only finitely many times and thus we can find a
corresponding accepting run of $\mA$: instead of each $\beta$-loop
that occurs before the last $\true$-loop we can use the $\true$-loop
since $\beta$ implies $\true$.

% When $\psi$ is not a temporal formula, this $\F$-merging would not
% save any state because there are no loops on states for conjunctions
% and disjunctions and thus such states are avoided already in the basic
% translation of $\F\psi$. However, we can still save some states in
% these cases if we further adjust the construction.

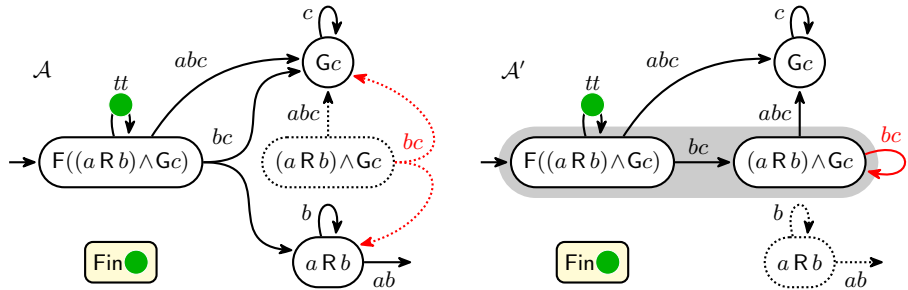
\begin{figure}[b]
\centering
\begin{tikzpicture}[automaton,transform shape,scale=.95]
\pgfdeclarelayer{background}
\pgfdeclarelayer{foreground}
\pgfsetlayers{background,main,foreground}
\def\RX{0}
\def\RY{-1.4}
\def\GX{\RX}
\def\GY{-\RY}
\def\andX{0}
\def\andY{0}
\def\FX{-2.9}
\def\FY{\andY}
\def\nameX{-4}
\def\nameY{1.3}
\def\labelX{\FX}
\def\labelY{\RY}
\def\boxPad{.5}
\begin{scope} []
  % % % The Ga state % % %
  \node[cstate] at (\GX,\GY) (Ga) {$\G c$};
  \path[->] (Ga) edge[loop above,outer sep=0pt] node[left,pos=.25] {$c$} (Ga);
  %
  % % % The Release state % % %
  \begin{scope}[]
  \node[cstate] (R) at (\RX,\RY) {$a\R b$};
  \path[->,auto]
  (R) edge[loop above] node[left,pos=.25] {$b$} (R)
      edge[overlay] node[below,pos=.4] {$ab$} + (1.2,0)
  ;
  \end{scope}
  %
  % % % The and-state % % %
  \begin{scope}[unreachable]
  \node[cstate] (RaG) at (\andX,\andY) {$(a\R b)\!\land\!\G c$};
  \path[->]
  (RaG) edge[] node[left] {$abc$} (Ga)
  ;
  \path[->,red]
    (RaG) edge[out=0,in=10,looseness=2]
          pic[above,pos=.1]{l=$bc$} (R.30)
    (RaG) edge[out=0,in=-10,looseness=2]  (Ga.-30)
  ;
  \end{scope}
  % % % The F-state % % %
  \begin{scope}
    \node[cstate,initial] at (\FX,\FY) (F) {$\F((a\R b)\!\land\!\G c)$};
    \path[->,auto]
      (F) edge[out=0,in=180,looseness=1.8] node[pos=.2] {$bc$} (Ga.-160)
      (F) edge[out=0,in=180,looseness=1.7] (R.160)
      (F.40) edge[bend left] node[] {$abc$} (Ga)
      (F) edge[loop above]
        pic {acc={3}{\phantom{T}}}
        pic {l=$\true$} (F)
    ;
  \end{scope}
  \node at (\nameX,\nameY) {$\mA$};
  \node[acclabel] at (\labelX,\labelY)
    {$\Fin\tacc{3}{\phantom{T}}$};
\end{scope}

\begin{scope}[xshift=6.6cm]
  % % % The Ga state % % %
  \node[cstate] at (\GX,\GY) (Ga) {$\G c$};
  \path[->] (Ga) edge[loop above,outer sep=0pt] node[left,pos=.25] {$c$} (Ga);
  %
  % % % The Release state % % %
  \begin{scope}[unreachable]
  \node[cstate] (R) at (\RX,\RY) {$a\R b$};
  \path[->,auto]
  (R) edge[loop above] node[left,pos=.25] {$b$} (R)
      edge[overlay] node[below,pos=.4] {$ab$} + (1.2,0)
  ;
  \end{scope}
  %
  % % % The and-state % % %
  \begin{scope}[]
  \node[cstate] (RaG) at (\andX,\andY) {$(a\R b)\!\land\!\G c$};
  \path[->]
  (RaG) edge[] node[left] {$abc$} (Ga)
  ;
  \path[->,red]
    (RaG.7) edge[out=20,in=-20,looseness=9] node[above,pos=.2]{$bc$} (RaG.-7)
  ;
  \end{scope}
  % % % The F-state % % %
  \begin{scope}
    \node[cstate,initial] at (\FX,\FY) (F) {$\F((a\R b)\!\land\!\G c)$};
    \path[->,auto]
      (F) edge node[pos=.4] {$bc$} (RaG)
      (F.40) edge[bend left] node[] {$abc$} (Ga)
      (F) edge[loop above]
        pic {acc={3}{\phantom{T}}}
        pic {l=$\true$} (F)
    ;
  \end{scope}
  \begin{pgfonlayer}{background}
  \draw[rounded corners=5mm,densely dashed,fill=black!20,draw=none] ($(F.180) + (-\boxPad/3,\boxPad)$)
     rectangle ($(RaG.0) + (\boxPad/3,-\boxPad)$);
  \end{pgfonlayer}
  \node at (\nameX,\nameY) {$\mA'$};
  \node[acclabel,thick] at (\labelX,\labelY)
    {$\Fin\tacc{3}{\phantom{T}}$};
\end{scope}
\end{tikzpicture}
\caption{Automata for $\F((a \R b)\land \G c)$: the SLAA $\mA$ built
  by the basic translation (left) and the modified SLAA $\mA'$ where
  states in the grey area can be merged (right).}
\label{fig:F-merging-and}
\end{figure}

The second example deals with the formula $\F\psi$ where $\psi=(a \R
b)\land\G c$. Figure~\ref{fig:F-merging-and} (left) depicts the SLAA
$\mA$ produced by the basic translation. %
The state $\psi$ is dotted as it is unreachable. Hence, merging
$\F\psi$ with $\psi$ would not save any state. However, we can modify
the translation rules to make $\psi$ reachable and $a \R b$
unreachable at the same time.  
% The automaton does not contain the state $\psi$ (as it is
% unreachable) and thus $\F\psi$ cannot be merged with it.
% Nevertheless,
%
The modification is based on the following observation. Taking the red
$bc$-edge in $\mA$ would mean that both $a \R b$ and $\G c$ have to
hold in the next step, which is equivalent to $(a \R b)\land\G c$.
Thus we can replace the red edge by the red $bc$-loop as shown in the
automaton $\mA'$ of Figure~\ref{fig:F-merging-and} (right). Because
transitions leaving the state $\F\psi$ are computed from transitions
of $\psi$, this replacement makes the state $(a \R b)\land\G c$
reachable and the state $a \R b$ becomes unreachable. The states
$\F\psi$ and $\psi$ of $\mA'$ can be now merged for the same reason as
in Figure~\ref{fig:f-merging:motivation}.

% % % modified AND % % %

\begin{figure}[t]
\centering
\begin{tikzpicture}[automaton]
\def\psiDist{2cm}
\def\psiAShift{0cm}
\def\psiBShift{1.cm}
\def\psiy{-.9}
\def\psiBy{-2.5}
\def\boxPad{.5cm}
\def\shorten{6pt}
\def\shortenOut{4pt}

\begin{scope}[unreachable]
	\node[cstate] (f) at (0,0) {$\F(\psi_1\lor\psi_2 \lor \psi_3)$};
	\node[state,fill=magenta!30] (p1) at (-\psiDist,\psiy) {$\psi_1$};
	\node[state,fill=orange!90!black!30] (p2) at (0,\psiy) {$\psi_2$};
	\node[state,fill=blue!50!cyan!30] (p3) at (\psiDist,\psiy) {$\psi_3$};
        \node[state,transparent] (ph) at (\psiDist,0) {$\psi$};
\end{scope}

\begin{scope}
  \node[state] (a1) at  (-\psiDist - \psiAShift,\psiBy) {};
  \node[state] (a2) at  (-\psiAShift,\psiBy) {};
  \node[state] (a3) at  (\psiDist - \psiAShift,\psiBy) {};
  \node[state] (b1) at  (-\psiDist + \psiBShift,\psiBy) {};
  \node[state] (b2) at  (\psiBShift,\psiBy) {};
  \node[state,overlay] (b3) at  (\psiDist + \psiBShift,\psiBy) {};
\end{scope}

\foreach \i/\ss/\a/\b/\c/\d in {1/\shorten - 2/0/3/2/2, 
                                2/\shorten/0/3/1/1,
                                3/\shorten/2/2/1/1}  {
  \foreach \s/\sh in {a/3,b/5.5} {
    \draw[->,shorten <=\sh] (p\i) -- (\s\i);
  }
  \draw (p\i.-90) edge[->,out=-90,in=-120,looseness=12,
                       shorten >=\ss,shorten <=\shorten] 
                       pic[pos=.2]{acc={\a}{\b}}
                        pic[pos=.7]{acc={\c}{\d}} (p\i);
}
    \draw[->,shorten <=2.9] (p3) -- (b3);

\draw[rounded corners=5mm,solid] (p1) + (-\boxPad,-\boxPad)
      rectangle (\psiDist + \boxPad,\boxPad);%($(p3) + (\boxPad,-\boxPad)$);
% \path[->]      
%  (f)  edge[loop above,looseness=11,
%  		  shorten >=\shorten,shorten <=\shortenOut]
%           pic{acc={3}{\phantom{0}}} pic{l=$\true$} (f)
%  ;
 \path[->]      
 (ph)  edge[loop right,looseness=11,
 		  shorten >=\shorten-1.5,shorten <=\shortenOut-1]
          pic{acc={3}{\phantom{0}}} pic{l=$\true$} (ph)
 ;

% \node[acclabel] at (5.5,-1.12) {$\Fin\gm\land(\Fin\tacc{1}{1} \lor \Fin\tacc{2}{2} \lor \Fin\tacc{0}{3})$};
\node[acclabel] at (0,-3.5) {$\Fin\gm\land(\Fin\tacc{1}{1} \lor \Fin\tacc{2}{2} \lor \Fin\tacc{0}{3})$};
\end{tikzpicture}
\caption{Transitions of the state $\F(\psi_1\vee\psi_2\vee\psi_3)$ merged with states $\psi_1$, $\psi_2$, and $\psi_3$.}
\label{fig:f-merging:or}
\end{figure}
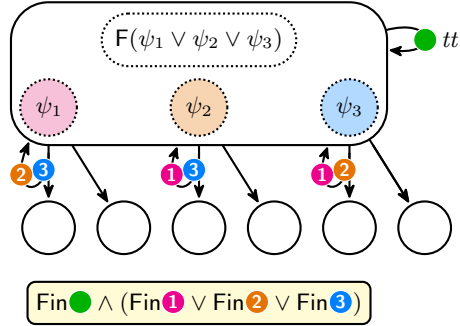

While the previous paragraph studied a formula $\F\psi$ where $\psi$
is a conjunction, the third example focuses on disjunctions. Let us
consider a formula $\F\psi$ where $\psi=\psi_1\vee\psi_2\vee\psi_3$
and each $\psi_i$ is a temporal formula. As in the previous example,
the state $\psi$ is unreachable in the SLAA produced by the basic
translation and thus merging $\F\psi$ with $\psi$ does not make any
sense. However, we can merge the state $\F\psi$ with states
$\psi_1,\psi_2,\psi_3$ as indicated in
Figure~\ref{fig:f-merging:or}. In contrast to the original SLAA, a
single run of the merged SLAA can use a loop corresponding to a state
$\psi_i$ and subsequently a transition corresponding to a different
$\psi_j$. Instead of every such a loop, the original SLAA can simply
use the $\true$-loop of $\F\psi$. However, as the $\true$-loop is
marked by $\gm$, we can use it only finitely many times. In fact, the
runs of the merged automaton that contain infinitely many loops
corresponding to two or more different states $\psi_i$ should be
nonaccepting. Therefore we adjust the acceptance formula to
$\Fin\gm\land(\Fin\tacc{1}{1} \lor \Fin\tacc{2}{2} \lor
\Fin\tacc{0}{3})$ and place the new acceptance marks as shown in
Figure~\ref{fig:f-merging:or}. Clearly, $\Fin\tacc{1}{1}$ says that
transitions of $\psi_2$ and $\psi_3$ are taken only finitely many
times, $\Fin\tacc{2}{2}$ does the same for transitions of $\psi_1$ and
$\psi_3$, and $\Fin\tacc{0}{3}$ for transitions of $\psi_1$ and
$\psi_2$.

The $\F$-merging translation combines the ideas presented
above when processing any $\F$-subformula, while other subformulae are
handled as in the basic translation. Hence, we no longer treat
$\F\psi$ as an abbreviation for $\true\U\psi$.
%
% Given an LTL formula
% $\varphi$, we define the sets of $\F$- and $\U$-subformulae as:
% \begin{align*}
% \mF_\varphi &= \{\F\psi \mid \F\psi\text{ is a subformula of } \varphi\}\\
% \mU_\varphi &= \{\psi_1\U\psi_2 \mid \psi_1\U\psi_2\text{ is a subformula of } \varphi\text{ and } \psi_1\neq\true\}
% \end{align*}
%
% We perform the $\F$-merging on formulae of the form
% $\F\bigvee_i\bigwedge_j\psi_{i,j}$ where $\psi_{i,j}$ are temporal
% formulae.
%
Further, we think about formulae $\F\psi$ as formulae of the form
$\F\bigvee_i\bigwedge_j\psi_{i,j}$, where $\psi_{i,j}$ are temporal
formulae. Formally, we define formula decomposition into disjunctive
normal form $\overline{\psi}$ as follows:
\begin{align*}
  \overline{\psi} &= \{\{\psi\}\}\textrm{ if }\psi\textrm{ is a temporal formula} \\
  \overline{\psi_1 \vee \psi_2} &= \overline{\psi_1}\cup\overline{\psi_2} \\
  \overline{\psi_1\land\psi_2} &= 
    \{C_1\cup C_2\mid C_1\in\overline{\psi_1}\textrm{ and }
                      C_2\in\overline{\psi_2}\}.
\end{align*}
Let $K\in\overline{\psi}$ be a set of temporal formulae. We use
$\psi_K$ to denote $\psi_K = \bigwedge_{\psi'\in K}\psi'$. Clearly,
$\psi$ is equivalent to $\bigvee_{K\in\overline{\psi}}\psi_K$. We
define two auxiliary transition functions
$\oDelta_\mathsf{L},\oDelta_\mathsf{NL}$ to implement the trick
illustrated with red edges in Figure~\ref{fig:F-merging-and}.
Intuitively, $\oDelta_\mathsf{L}(\psi_K,\alpha)$ is the set of
$\alpha$-transitions of $\psi_K$ such that their destination
configuration subsumes $K$ (the transitions are looping in this sense,
hence the subscript $\mathsf{L}$), while
$\oDelta_\mathsf{NL}(\psi_K,\alpha)$ represents the remaining
$\alpha$-transitions of $\psi_K$ (non-looping, hence the subscript
$\mathsf{NL}$). To mimic the trick of Figure~\ref{fig:F-merging-and},
we should replace in the destination configuration of each looping
transition all elements of $K$ by the state corresponding to
$\psi_K$. To simplify this step, we define the destination
configurations of looping transitions in
$\oDelta_{\mathsf{L}}(\psi_K,\alpha)$ directly without elements of
$K$.
\begin{align*}
\oDelta_{\mathsf{L}}(\psi_K,\alpha)  &= \big\{(M,C\smallsetminus K) \mid 
                                      (M,C) \in \Delta(\psi_K,\alpha),\, 
                                      K\subseteq C \big\}\\
\oDelta_{\mathsf{NL}}(\psi_K,\alpha) &= \big\{(M,C)\mid 
                                      (M,C) \in \Delta(\psi_K,\alpha),\,
                                      K\not\subseteq C\big\}
\end{align*}
Simultaneously with the trick of Figure~\ref{fig:F-merging-and}, we
apply the idea indicated in Figure~\ref{fig:f-merging:or} and merge
the state $\F\psi$ with all states $\psi_K$ for
$K\in\overline{\psi}$. Hence, instead of extending the looping
transitions with states $\psi_K$, we extend it with the merged state
called simply $\F\psi$. Altogether, we get
\begin{align*}
\Delta(\F\psi,\alpha) = {}&\big\{(\{\gm\},\{\F\psi\})\big\}
  \cup {} \\
  &\cup \bigcup_{K\in\overline{\psi}}
    \Big(\dm\big(\oDelta_{\mathsf{NL}}(\psi_K,\alpha)\big)
    ~\cup~ \big\{(M_K,\{\F\psi\})\big\}
    \otimes\oDelta_\mathsf{L}(\psi_K,\alpha)\Big)
  \text{ where} \\M_K = {}&\{\om^{K'}\mid 
         K'\in\overline{\psi}\text{ and } K'\neq K \}.
\end{align*}
In other words, the merged state $\F\psi$ has three kinds of
transitions: the $\true$-loop marked by \gm, non-looping transitions
of states $\psi_K$ for each disjunct $K\in\overline{\psi}$, and the
looping transitions of states $\psi_K$ which are marked as shown in
Figure~\ref{fig:f-merging:or}.
Finally, we redefine the set of acceptance marks $\mM$ and the
acceptance formula~$\varPhi$ of the constructed SLAA as follows, where
$\mF_\varphi$ denotes the set of all subformulae of $\varphi$ of the
form $\F\psi$:
\begin{align*}
\mM &= \{\gm\} \cup \{\om^K\mid \F\psi\in\mF_\varphi \text{ and } K\in\overline{\psi} \}\\
\varPhi &= \Fin\gm ~\land
\bigwedge_{\F\psi\in\mF_\varphi\!\phantom{\overline{\psi}}}\!\!
\bigvee_{~K\in\overline{\psi}}\!\Fin\om^K
\end{align*}
% \[
%   \mM = \{\gm\} \cup \{\om^K\mid \F\psi\in\mF_\varphi \text{ and } K\in\overline{\psi} \}
%   \hspace{8ex}
%   \varPhi = \Fin\gm ~\land \!\bigwedge_{\F\psi\in\mF_\varphi\!\phantom{\overline{\psi}}}\!\!\! \bigvee_{~K\in\overline{\psi}}\!\Fin\om^K
% \]

In fact, we can reduce the number of orange marks (marks with an upper
index) produced by $\F$-merging translation. Let
$n_\varphi=\max\{|\overline{\psi}|\mid\F\psi\in\mF_\varphi\}$ be the
maximal number of such marks corresponding to any subformula $\F\psi$
of $\varphi$. We can limit the total number of orange marks to
$n_\varphi$ by reusing them. We redefine $\mM$ and $\Phi$ as 
\[
  \mM = \{\gm\} \cup \{\om^i\mid 1 \leq i \leq n_\varphi \} \textrm{~~~~~~and~~~~~~} \Phi = \Fin\gm \land \bigvee_{i=1}^{n_\varphi}\Fin\om^i
\]
and alter the above definition of $M_K$. For every
$\F\psi\in\mF_\varphi$, we assign a unique index $i_K$,
$1\le i_K\le n_\varphi$, to each $K\in\overline\psi$.
% ($i_K$ can differ for two occurrences of $K$ in distinct $\F\psi_1, \F\psi_2$).
Sets $M_K$ in transitions of $\F\psi$ are then defined to contain all
orange marks except $\om^{i_K}$, formally
$M_K=\mM\smallsetminus\{\gm,\om^{i_K}\}$. This optimization is correct
as any branch of an SLAA run cannot cycle between two states of the form $\F\psi$.

\section{$\FG$-Merging Translation}
\label{sec:fg-merging}

We further improve the $\F$-merging translation by adding a special
rule also for subformulae of the form $\G\psi$. The resulting
$\FG$-merging translation produces SLAA with an acceptance formula
that is not $\Inf$-less.
% However, it still maintains a specific form
% that is suitable for the efficient dealternation algorithm presented
% by Major~\cite{major.17.thesis} and implemented in the tool LTL3TELA\todo{cite toolpaper}.

% \[\mO(\gm) =
%   \{s\in S\mid (s,\alpha,M,C) \in \Delta \text{ and } \gm\in M\}\]
%An SLAA is called a \emph{single-owner SLAA} if $\mO(\gm)$ is a
%singleton for each acceptance mark $\gm$. A \emph{sibling SLAA} is a
%single-owner SLAA with an acceptance formula $\varPsi$ where each
%$\Inf\bsq$ term occurs only in disjunction with some $\Inf$-less
%formula $\varPsi'$ different from $\false$ and such that all marks
%appearing in $\varPsi'$ have the same owner as $\bsq$.

% In this section we present a class of SLAA that can be more succinct than
% $\Inf$-less SLAA on one hand and can be still efficiently dealternated on the
% other hand.  Let $\mA$ be a single-owner SLAA with an acceptance formula
% $\varPsi$. We say that $\mA$ is a \emph{sibling automaton} if it holds that any
% $\Inf\bsq$ term in $\varPsi$ occurs only in disjunction with some $\Inf$-less
% formula $\varPsi'$ such that all marks that appear in $\varPsi'$ share the owner
% with $\bsq$.

\begin{figure}[t]
\centering
\begin{tikzpicture}[semithick,>={Stealth[round,bend]},xscale=.82]
\def\length{6}
\def\Tlength{5}
\def\ySeg{-1.9}
\def\xShiftT{.5}
\def\xbase{2.5}
\def\yGF{0}
\def\yFa{-1.3}
\def\labelpos{.32}

\begin{scope}
% % % STATES % % %
\begin{scope}[automaton,transform shape=false,xscale=1]
\node[cstate,initial,initial angle=90,initial overlay] (GF) at (0,\yGF) {$\G\F a$};
\node[cstate] (Fa) at (0,\yFa) {$\F a$};

% % % automata transitions % % %
\path[->,auto]
(GF) edge node {$\true$} (Fa)
(GF.-90) edge[out=-90,in=-120,looseness=6] (GF)
(GF) edge [loop right] node {$a$} (GF)
(Fa) edge [loop left]
  pic {acc={3}{\phantom{0}}} pic {l=$\true$} (Fa)
(Fa) edge node {$a$} + (1.2,0)
;

% % % Name tag & acclabel % % %
\node[anchor=east] at (-2.2,0) {$\mA$};
\node[acclabel,anchor=east] at (-2.2,-.65) {$\Fin\gm$};
\end{scope}

% % % SEGMENTS % % %
\path[dotted]
foreach \x in {0,...,\length} {(\xbase+\x,\yGF+.2) edge (\xbase+\x,\ySeg)}
foreach \x in {0,...,\length} {node[above,font=\tiny] at(\x+\xbase,.2) {\x}}
%foreach \x in {0,...,\Tlength} {node[below] at (\xbase+\x+\xShiftT,\ySeg) {$T_\x$}}
;

% % % ... % % %
\node at (\xbase+\length+\xShiftT+.3,-1) {$\cdots$};

% % % GF dots % % %
\foreach \x in {0,...,\length} \node[dot] (GF\x) at (\xbase+\x,\yGF) {};
\foreach \x in {0,...,\length} \node[dot,transparent,overlay] (GF\x+1) at (\xbase+\x+1,\yGF) {};
\foreach \x in {0,...,\length} \node[dot,transparent,overlay] (GF\x+2) at (\xbase+\x+2,\yGF) {};

% % % Fa dots % % %
\foreach \x in {1,2,4,5} \node[dot] (Fa\x) at (\xbase+\x,\yFa) {};
\foreach \x in {0,...,\length} \node[dot,transparent,overlay] (Fa\x+1) at (\xbase+\x+1,\yFa) {};
\foreach \x in {0,...,\length} \node[dot,transparent,overlay] (Fa\x-1) at (\xbase+\x-1,\yFa) {};

% % % edges % % %
\path[->]
\foreach \i in {0,1,3,4} {
(GF\i) edge node[above=.15] {$\emptyset$} (GF\i+1)
(GF\i) edge (Fa\i+1)
}
\foreach \i in {2,5} {
(GF\i) edge node[above=.15] {$\{a\}$} (GF\i+1)
(Fa\i-1) edge node[pos=.45]{\gm} (Fa\i)
{(Fa\i) edge +(.7,-.5)}
}
;
\end{scope}

\begin{scope}[yshift=-3.5cm]
\def\ySeg{-.3}
% % % STATES % % %
\begin{scope}[automaton,transform shape=false,xscale=1]
\node[cstate,initial,initial angle=90] (GF) at (0,\yGF) {$\G\F a\land \F a$};

% % % automata transitions % % %
\path[->,auto]
(GF.189) edge [out=195,in=165,looseness=7]
  pic {acc={3}{\phantom{0}}} pic[left] {l=$\true$} (GF.171)
(GF.9) edge [in=-15,out=15,looseness=7]
  pic[] {accsq={0}{\phantom{0}}} pic {l=$a$} (GF.-9)
;

\node[anchor=east] at (-2.2,.65) {$\mA_{\FG}$};
\node[acclabel,anchor=east] at (-2.2,0) {$\Fin\gm\lor\Inf\bsq$};
\end{scope}

% % % SEGMENTS % % %
\path[dotted]
foreach \x in {0,...,\length} {(\xbase+\x,\yGF+.3) edge (\xbase+\x,\ySeg)}
;

% % % ... % % %
\node at (\xbase+\length+\xShiftT+.3,0) {$\cdots$};

% % % GF dots % % %
\foreach \x in {0,...,\length} \node[dot] (GF\x) at (\xbase+\x,\yGF) {};
\foreach \x in {0,...,\length} \node[dot,transparent,overlay] (GF\x+1) at (\xbase+\x+1,\yGF) {};
\foreach \x in {0,...,\length} \node[dot,transparent,overlay] (GF\x+2) at (\xbase+\x+2,\yGF) {};

% % % edges % % %
\path[->]
\foreach \i in {0,3} {
(GF\i) edge (GF\i+1)
(GF\i+1) edge
  pic[pos=.45]{acc={3}{\phantom{0}}} (GF\i+2)
}
\foreach \i in {2,5} {
(GF\i) edge
  pic[pos=.45]{accsq={0}{\phantom{0}}} (GF\i+1)
}
\foreach \i in {2,5} {
%(GF\i) edge node[above=.15] {$\{a\}$} (GF\i+1)
}
;
\end{scope}
\end{tikzpicture}
\caption{An SLAA $\mA$ for the formula $\G\F a$ built by the basic
translation (top) and an equivalent SLAA $\mA_{\FG}$ built by
the $\FG$-merging translation (bottom) and their runs over the word
$(\emptyset\emptyset\{a\})^\omega$.} \label{fig:fg-merging:motivation}
\end{figure}
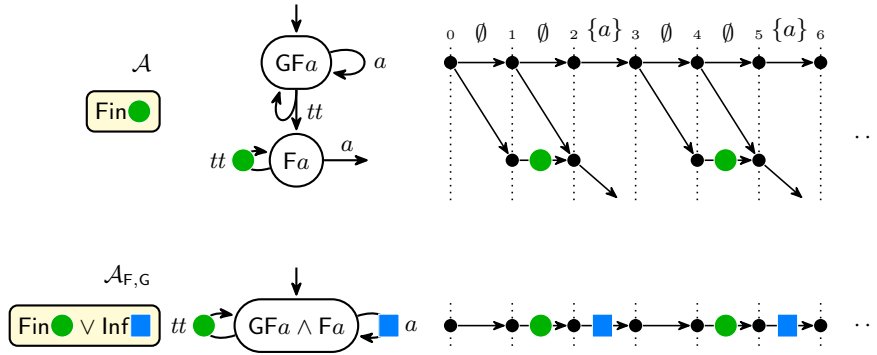

We start again with a simple example. Consider the formula $\G\F a$.
The basic translation produces the SLAA $\mA$ from
Figure~\ref{fig:fg-merging:motivation} (top). In general, each
transition of the state $\G\psi$ is a transition of $\psi$ extended
with a loop back to $\G\psi$. The one-to-one correspondence between
transitions of $\G\psi$ and $\psi$ leads to the idea to merge the
states into one that corresponds to their conjunction
$(\G\psi)\land\psi$, which is equivalent to $\G\psi$. However, merging
these states needs a special care.
Figure~\ref{fig:fg-merging:motivation} (bottom) shows an SLAA where
the states $\G\F a$ and $\F a$ are merged. Consider now the word $u =
(\emptyset\emptyset\{a\})^\omega$ and the runs of the two automata
over $u$. The branches of the top run collapse into a single branch in
the bottom run. While each branch of the top run has at most one
occurrence of $\gm$, the single branch in the bottom run contains
infinitely many of these marks. The SLAA $\mA_{\FG}$ accepts $u$ only
because of the added $\bsq$ marks. The intuition for their placement
is explained using the concept of \emph{escaping multitransitions}.

A multitransition $T$ of an SLAA $\mA'$ is \emph{$s$-escaping} for a
state $s$ if it contains a non-looping transition $(s,\alpha,M,C)\in
T$. For an acceptance mark $\gm$, we define its \emph{owners}
$\mO(\gm)=\{s\in S\mid (s,\alpha,M,C) \in \Delta \text{ and } \gm\in
M\}$ as the set of all states with outgoing transitions marked by \gm.
The following observation holds for every mark $\gm$ with a single
owner $s$. \emph{A run of $\mA'$ satisfies $\Fin\gm$ (i.e., all its
infinite branches satisfy $\Fin\gm$) if and only if the run contains
only a finite number of multitransitions $T$ marked by $\gm$ or if it
contains infinitely many $s$-escaping multitransitions.}

Transitions of $\mA_{\FG}$ correspond to multitransitions of $\mA$
with source configuration $\{\G\F a,\F a\}$. The observation implies
that $\mA_{\FG}$ would be equivalent to $\mA$ if we mark all
transitions corresponding to $\F a$-escaping multitransitions with a
new mark $\bsq$ and change the acceptance formula to
$\Fin\gm\vee\Inf\bsq$ as shown in
Figure~\ref{fig:fg-merging:motivation} (bottom).

% Consider a formula $\G\psi$ where $\psi$ is a temporal formula, the
% corresponding SLAA $\mA_\F$ build by the $\F$-merging construction is depicted
% in Figure~\ref{fig:fg-merging:motivation} (we take the fact that $\psi$ can be an $\F$ or $\U$ formula into account).  For each transition $t$ of $\psi$
% the state $\G\psi$ has a transition that follows the destinations of $t$ and
% loops back to $\G\psi$ on top of that. Therefore, the states can be merged into
% a state that represent their conjunction (which is equivalent to $\G\psi$) where
% each transition corresponds to a multitransition that consists of two
% transitions: one for $\G\psi$ and one for $\psi$.  However all branches of runs
% of $\mA_\F$ that stay in $\G\psi$ or $\psi$ will be merged into one, by this
% merging.  This situation resemble the problem we had to solved for dealternation
% of $\Inf$-less SLAA.  Indeed, also the solution is inspired from there.  We want
% to follow $\psi$-escaping multitransitions, therefore we mark all transitions
% that correspond to a non-looping transition of $\psi$ by $\bsq_\psi$.  The
% merged state for $\G\psi$ is depicted as $\mA_{\FG}$ in
% Figure~\ref{fig:fg-merging:motivation} (right).

This approach naturally extends to the case of $\G\psi$ where
$\psi=\bigwedge_i\psi_i$ is a conjunction of temporal formulae. In
this case, $\G\psi$ can be merged with all states $\psi_i$ into a
single state representing the conjunction
$(\G\psi)\land\bigwedge_i\psi_i$. However, we have to pay a special
attention to acceptance marks as the observation formulated above
holds only for marks with a single owner. For every $\psi_i$ with a
$\gm$-marked transition, we need to track $\psi_i$-escaping
multitransitions separately. To implement this, we create a copy of
$\gm$ for each $\psi_i$ and we use a specific $\bsq_{\psi_i}$
mark to label transitions corresponding to $\psi_i$-escaping
multitransitions.

Unfortunately, due to the duality of the $\F$ and $\G$ operators and
the fact that the transition relation of SLAA is naturally in
disjunctive normal form (which is suitable for $\F$), we did not find
any way to improve the translation of $\G\psi$ if $\psi$ is a
disjunction. On the bright side, we can generalize the merging to
$\G\bigwedge_i\psi_i$ where each $\psi_i$ is a temporal or state
formula (which can contain disjunctions). This is due to the fact that
a state formula $\psi_i$ affects only the labels of transitions with
origin in the state $\G\bigwedge_i\psi_i$ and thus it does not create
any reachable state or acceptance mark.

Formally, we first modify the translation rules introducing acceptance
marks to work with marks of the form $\gm_{\psi}$ as discussed
above. More precisely, we change the rule for $\psi_1 {\U} \psi_2$
presented in the basic translation and the rule for $\F\psi$ of the
$\F$-merging translation to the following (note that the optimization reusing orange marks make no longer sense as we need to create their copies for each $\F\psi$ anyway).
\begin{align*}
\Delta(\psi_1 {\U} \psi_2,\alpha) = {}&
     \dm\big(\Delta(\psi_2,\alpha)\big) \!\cup\!
     \Big(\!\big\{(\{\gm_{\psi_1 {\U} \psi_2}\}, \{\psi_1 {\U} \psi_2\})\big\} \!\otimes\!
     \dm\big(\Delta(\psi_1,\alpha)\big)\!\Big)\\
\Delta(\F\psi,\alpha) = {}&\big\{(\{\gm_{\F\psi}\},\{\F\psi\})\big\}
  \cup{}\\
  &\cup \bigcup_{K\in\overline{\psi}}
    \Big(\dm\big(\oDelta_{\mathsf{NL}}(\psi_K,\alpha)\big)
    \cup \big\{(M_K,\{\F\psi\})\big\}
    \otimes\oDelta_\mathsf{L}(\psi_K,\alpha)\Big),\\
    \text{where }M_K = {}&\{\om_{\F\psi}^{K'}\mid 
         K'\in\overline{\psi}\text{ and } K'\neq K \}
\end{align*}

Further, we add a specific rule for formulae $\G\psi$ where
$\psi=\bigwedge_{\psi'\in K}\psi'$ for some set $K$ of temporal and
state formulae. Formulae $\G\psi$ of other forms are handled as
$\false\R\psi$.
On the top level,
the rule simply defines transitions of $\G\psi$ as transitions of
$\G\psi~\wedge\bigwedge_{\psi'\in K}\psi'$:
\[
\Delta(\G\psi,\alpha) =
    \big\{(\emptyset,\{\G\psi\}) \big\} \otimes
    \bigotimes_{\psi'\in K} \Delta'(\psi',\alpha)
\]
The definition of $\Delta'(\psi',\alpha)$ differs from
$\Delta(\psi',\alpha)$ in two aspects. First, it removes $\psi'$ from
destination configurations because $\psi'$ is merged with $\G\psi$ and
the state $\G\psi$ is added to each destination configuration by the
product on the top level. Second, it identifies all non-looping
transitions of $\psi'$ and marks them with $\bsq_{\psi'}$ as
$\psi'$-escaping.  We distinguish between looping and non-looping
transitions only when $\psi'$ has the form $\psi_1\U\psi_2$ or
$\F\psi_1$. All other $\psi'$ have only looping transitions
(e.g.,~$\G$-formulae) or no marked transitions (e.g.,~state formulae
or $\R$- or $\X$-formulae) and thus there are no $\psi'$-escaping
transitions or we do not need to watch them. Similarly to
$\mF_\varphi$, we use $\mU_\varphi$ for the set of all subformulae of
$\varphi$ of the form $\psi_1\U\psi_2$. The function
$\Delta'(\psi',\alpha)$ is defined as follows:
\begin{align*}
  \Delta'(\psi',\alpha) = {}&
  \begin{cases}
    \Delta'_\mathsf{L}(\psi',\alpha) \cup \Delta'_\mathsf{NL}(\psi',\alpha)
      &\text{if }\psi' \in \mF_\varphi\text{ or }\psi'\in\mU_\varphi\\[1.5ex]
    \big\{(M,C\smallsetminus\{\psi'\}) \mid (M,C) \in \Delta(\psi',\alpha)\big\} &\text{otherwise}
  \end{cases}\\[2ex]
\Delta'_\mathsf{L}(\psi',\alpha) = {}& \big\{(M,C\smallsetminus\{\psi'\})\mid (M,C) \in\Delta(\psi',\alpha), \psi'\in C \big\} \\
\Delta'_\mathsf{NL}(\psi',\alpha) = {}& \big\{(\{\bsq_{\psi'} \},C)\mid (M,C) \in\Delta(\psi',\alpha), \psi'\notin C \big\}
\end{align*}

Finally, we redefine the set of marks $\mM$ and the acceptance formula $\Phi$. Now each subformula from $\mU_\varphi$ and $\mF_\varphi$ has its own set of marks, and the $\bsq$ marks are used to implement the intuition given using the Figure~\ref{fig:fg-merging:motivation}.
% \begin{align*}
%    \mM &= \mM_\F \cup \mM_{\U{}} \text{ where}
%    \\ \mM_\F &=\{\gm_{\F\psi}, \bsq_{\F\psi},\om_{\F\psi}^K\mid
%        \F\psi\in\mF_\varphi \text{ and } K\in\overline{\psi} \}\\
%        \mM_{\U{}} &=\{\gm_{\psi},\bsq_{\psi} \mid \psi\in\mU_\varphi\}\\
%    \varPhi &= \bigwedge_{\psi\in\mU_\varphi}\!\!\left(\Fin\gm_{\psi} \lor \Inf\bsq_{\psi}\right) \,\land
%    \bigwedge_{\F\psi\in\mF_\varphi}\!\!\!\!\varPhi_{\F\psi}\text{ where}\\
%    \varPhi_{\F\psi} &= \left(\Fin\gm_{\F\psi} \land
%    \bigvee_{K\in\overline{\psi}}\!\!\Fin\om_{\F\psi}^K\right) \lor \Inf\bsq_{\F\psi}
% \varPhi' &= \bigwedge_{\substack{\psi\in\mU_\varphi\\s\in\mO(\psi)}}\!\!\left(\Fin\gm_{\psi,s} \lor \Inf\bsq_{\psi,s}\right) \,\land
%    \bigwedge_{\substack{\F\psi\in\mF_\varphi\\s\in\mO(\F\psi)}}\!\!\!\!\varPhi_{\F\psi,s}\text{ where}\\
% \varPhi_{\F\psi,s} &= \left(\Fin\gm_{\F\psi,s} \land
%    \bigvee_{K\in\overline{\psi}}\!\!\Fin\om_{\F\psi,s}^K\right) \lor \Inf\bsq_{\F\psi,s}
% \end{align*}
\begin{align*}
\mM &=
    %Untils
    \{\gm_{\psi},\bsq_{\psi} \mid \psi\in\mU_\varphi\}
  \cup
    \left\{\gm_{\F\psi}, \bsq_{\F\psi},\om_{\F\psi}^K\mid
    \F\psi\in\mF_\varphi \text{ and } K\in\overline{\psi} \right\}
  \\
\Phi &=\!\!\!\!
\bigwedge_%
  {\psi\in\mU_\varphi}
  \!\!\!
  \left(\Fin\gm_{\psi} \lor \Inf\bsq_{\psi}\right) \,\land\!\!\!
\bigwedge_%
  {\F\psi\in\mF_\varphi}
  \!\!
  \left(\Big(\Fin\gm_{\F\psi} \land\!\!
   \bigvee_{K\in\overline{\psi}}\!\!\Fin\om_{\F\psi}^K\Big) \lor \Inf\bsq_{\F\psi}\right)
\end{align*}

% Each transition of $\G\psi$ is a loop ($P_0$). The loops of states for $\psi'\in
% K$ are redirected to $\G\psi$ ($C\smallsetminus\{\psi'\}$ in combination with
% $P_0$).  The division of $P_{\psi'}$ to $P_\mathsf{L}^{\psi'}$ and
% $P_\mathsf{NL}^{\psi'}$ follows the ideas of the modified translation for
% conjunctions from the previous section.  The looping transitions are just
% redirected to $\G\psi$ ($P_\mathsf{L}^{\psi'}$) while the non-looping
% transitions are marked by $\bsq_{\psi'}$ ($P_\mathsf{NL}^{\psi'}$).

\begin{theorem}\label{thm:fg-merging}
  Let $\varphi$ be an LTL formula and let $\mA_\varphi$ be the corresponding SLAA built by the $\FG$-merging translation. Then $L(\mA_\varphi) = L(\varphi)$.
  Moreover, the number of states of $\mA_{\varphi}$ is linear to the size of $\varphi$
  and the number of acceptance marks is at most exponential to the size of $\varphi$.
\end{theorem}
The proof can be found in
\iffullversion
Appendix~\ref{app:proof}.
\else
the full version of this paper~\cite{blahoudek.19.arxiv}.
\fi

%}}}
%{{{ SLAA simplification

\section{SLAA Simplification}
\label{sec:simplification}

Simplification by transition dominance is a basic technique improving
various automata constructions~\cite{gastin.01.cav}.  The idea is that
an automata construction does not add any transition that is dominated
by some transition already present in the automaton.  In SLAA, a
transition $(q,\alpha,M_1,C_1)$ \emph{dominates} a transition
$(q,\alpha,M_2,C_2)$ if $C_1\subseteq C_2$ and $M_1$ is ``at least
as helpful and at most as harmful for acceptance'' as $M_2$. In the
rest of this section, we focus on the precise formulation of the last
condition.

In the classic case of co-Büchi SLAA with acceptance formula
$\Fin\gm$, the condition translates into
$\gm\in M_1\implies \gm\in M_2$.  For Büchi SLAA with acceptance
formula $\Inf\bsq$, the condition has the form
$\bsq\in M_2\implies \bsq\in M_1$.

Now consider an SLAA with an arbitrary acceptance formula $\Phi$. Let
$\Fin(\Phi)$ and $\Inf(\Phi)$ be the sets of all acceptance marks
appearing in $\Phi$ in subformulae of the form $\Fin\gm$ and
$\Inf\bsq$, respectively. A straightforward formulation of the
condition is $M_1\cap\Fin(\Phi)\subseteq M_2$ and
$M_2\cap\Inf(\Phi)\subseteq M_1$. This formulation is correct, but we
can do better. For example, consider the case
$\Phi=\Fin\tacc{3}{1}\wedge(\Fin\tacc{2}{2}\vee\Inf\taccsq{1}{3})$ and
transitions $t_1=(q,\alpha,\{\tacc{3}{1},\tacc{2}{2}\},\{p\})$ and
$t_2=(q,\alpha,\{\tacc{3}{1}\},\{p,p'\})$.  Then $t_1$ does not
dominate $t_2$ according to this formulation as
$\{\tacc{3}{1},\tacc{2}{2}\}\cap\Fin(\Phi)=\{\tacc{3}{1},\tacc{2}{2}\}\not\subseteq\{\tacc{3}{1}\}$.
However, any branch of an accepting run cannot take neither $t_1$ nor
$t_2$ infinitely often and thus $t_1$ can be seen as dominating in this
case. To formalize this observation, we introduce \emph{transition
  dominance with respect to acceptance formula}.

A \emph{minimal model} $O$ of an acceptance formula $\Phi$ is a subset
of its terms satisfying $\Phi$ and such that no proper subset of $O$
is a model of $\Phi$. For example, the formula
$\varPhi=\Fin\tacc{3}{1} \wedge (\Fin\tacc{2}{2} \vee
\Inf\taccsq{1}{3})$ has two minimal models:
$\{\Fin\tacc{3}{1},\Fin\tacc{2}{2}\}$ and
$\{\Fin\tacc{3}{1},\Inf\taccsq{1}{3}\}$. For each minimal model $O$,
by $\Fin(O)$ and $\Inf(O)$ we denote the sets of all acceptance marks
appearing in $O$ in terms of the form $\Fin\gm$ and $\Inf\bsq$,
respectively. We say that a transition $(q,\alpha,M_1,C_1)$
\emph{dominates} a transition $(q,\alpha,M_2,C_2)$ \emph{with respect
  to} $\Phi$ if $C_1\subseteq C_2$ and for each minimal model $O$
of $\Phi$ it holds
$\Fin(O) \cap M_2 = \emptyset \implies \Fin(O) \cap M_1 = \emptyset$
and
$\Inf(O) \cap M_1 = \emptyset \implies \Inf(O) \cap M_2 = \emptyset$.
In other words, if $t_2$ can be used infinitely often without breaking
some $\Fin\gm$ of a minimal model then $t_1$ can be used as well, and
if an infinite use of $t_1$ does not satisfy any term $\Inf\bsq$ of a
minimal model then an infinite use of $t_2$ does not as well.

Note that our implementation of LTL to SLAA translation used later in
experiments employs this simplification.

%}}}
%{{{ translation of SLAA to LTL

\section{Translation of SLAA to LTL}
\label{sec:slaa2ltl}

This section presents a translation of an SLAA to an equivalent LTL
formula. Let $\mA=(S,\Sigma,\mM,\Delta,s_I,\varPhi)$ be an SLAA. We
assume that the alphabet has the form $\Sigma=2^{\AP'}$ for some
finite set of atomic propositions $\AP'$. For each $\alpha\in\Sigma$,
$\varphi_\alpha$ denotes the formula
\[
  \varphi_\alpha=\bigwedge_{a\in\alpha}a~~~\wedge~~~\bigwedge_{\mathclap{a\in\AP'\smallsetminus\{\alpha\}}}\neg a.
\]    
For each state $s\in S$ we
construct an LTL formula $\varphi(s)$ such that $u\in\Sigma^\omega$
satisfies $\varphi(s)$ if and only if $u$ is accepted by $\mA$ with
its initial state replaced by $s$. Hence, $\mA$ is equivalent to the
formula $\varphi(s_I)$.

The formula construction is inductive. Let $s$ be a state such that we
have already constructed LTL formulae for all its successors. Further,
let $\Mod(\varPhi)$ denote the set of all minimal models of $\varPhi$.
Further, given a set of states $C$, by $\varphi(C)$ we denote the
conjunction $\varphi(C)=\bigwedge_{s\in C}\varphi(s)$.
We define $\varphi(s)$ as follows:
\[
  \begin{array}{rcl}
  \varphi(s) & = & \varphi_1(s) \vee \big(\varphi_2(s)\wedge\varphi_3(s)\big)\\[3ex]
  \varphi_1(s) & = & \displaystyle
    \bigvee_{\substack{(s,\alpha,M,C)\in\Delta\\s\in C}}\!\!\!\!\!\!\!\!
    \varphi_\alpha\wedge\X\varphi(C\smallsetminus\{s\})~~~~~\U
    \bigvee_{\substack{(s,\alpha,M,C)\in\Delta\\s\not\in C}}\!\!\!\!\!\!\!\! 
    \varphi_\alpha\wedge\X\varphi(C)\\[6ex]
  \varphi_2(s) & = & \displaystyle 
    ~\,\G \!\!\!\!\!\!\!\!\bigvee_{\substack{(s,\alpha,M,C)\in\Delta\\s\in C}}\!\!\!\!\!\!\!\! 
    \varphi_\alpha\wedge\X\varphi(C\smallsetminus\{s\})\\[6ex]
  \varphi_3(s) & = & \displaystyle 
    \bigvee_{O\in\Mod(\varPhi)}\Big(
    ~~~~\bigwedge_{\bsq\in \Inf(O)}
    ~\big(~\G\F\!\!\!\!\!\!\!\bigvee_{\substack{(s,\alpha,M,C)\in\Delta\\s\in C,~\bsq\in M}}
    \!\!\!\!\!\!\!\!\varphi_\alpha\wedge\X\varphi(C\smallsetminus\{s\})~\big) 
    ~~\wedge\\[6ex]
  && \displaystyle ~~~~~~~~~~~~~~\,\wedge
    \bigwedge_{\gm\in \Fin(O)}
    ~\big(~\F\G\!\!\!\!\!\!\!\bigvee_{\substack{(s,\alpha,M,C)\in\Delta\\s\in C,~\gm\not\in M}}
    \!\!\!\!\!\!\!\!\varphi_\alpha\wedge\X\varphi(C\smallsetminus\{s\})~\big)~~\Big) 
  \end{array}
\]
Intuitively, $\varphi_1(s)$ covers the case when a run leaves $s$
after a finite number of looping transitions. Note that $\varphi_1(s)$
ignores acceptance marks on looping transitions of $s$ as they play no
role in acceptance of such runs. Further, $\varphi_2(s)$ describes the
case when a run never leaves $s$. In this case, $\varphi_3(s)$ ensures
that the infinite branch of the run staying in $s$ forever is
accepting. Indeed, $\varphi_3(s)$ says that the branch satisfies some
prime implicant of the acceptance condition $\varPhi$ as all
$\Inf$-marks of the prime implicant have to appear infinitely often on
the branch and each $\Fin$-mark of the prime implicant does not appear at
all after a finite number of transitions.

%\medskip The following theorem states correctness of the translation.
\begin{theorem}\label{thm:slaa2ltl}
  Let $\mA$ be an SLAA over an alphabet $\Sigma=2^{\AP'}$ and with initial
  state $s_I$. Further, let $\varphi(s_I)$ be the formula constructed
  as above. Then $L(\mA)=L(\varphi(s_I))$.
\end{theorem}
The statement can be proven straightforwardly by induction reflecting
the inductive definition of the translation.

Theorems~\ref{thm:fg-merging} and~\ref{thm:slaa2ltl} imply that LTL
and the class of SLAA with alphabets of the form $2^{\AP'}$ have the
same expressive power.

%}}}
%{{{ experiments

\section{Implementation and Experimental Evaluation}%
\label{sec:experiments}%

We have implemented the presented translations in a tool called
\emph{LTL3TELA} which also offers dealternation algorithms for
$\Inf$-less SLAA and SLAA produced by the $\FG$-merging translation.
Given an LTL formula, LTL3TELA first applies the formula optimization
process implemented in SPOT. The processed formula is then translated
to SLAA by one of the three presented translation, unreachable states are removed and transition reductions suggested by Babiak et
al.~\cite{babiak.12.tacas} are applied.

\subsection{Evaluation Settings}

We compare the LTL to SLAA translation implemented in LTL3BA and the
translations presented in this paper as implemented in LTL3TELA.
Table~\ref{tab:tools} provides homepages and versions numbers of SPOT
and both compared translators.
%Table~\ref{tab:commands} shows the
%commands for executing the tools.
All scripts and formulae used for
the evaluation presented below are available in a Jupyter notebook that can be found at 
\texttt{\href{https://github.com/jurajmajor/ltl3tela/blob/master/Experiments/Evaluation_ICTAC19.ipynb}{https://github.com/jurajmajor/}
\href{https://github.com/jurajmajor/ltl3tela/blob/master/Experiments/Evaluation_ICTAC19.ipynb}{ltl3tela/blob/master/Experiments/Evaluation\_ICTAC19.ipynb}}.

The
presented improvements can only be applied on certain kinds of
formulae: formulae that contain at least one $\F\psi$ subformula where
$\psi$ contains some temporal subformula or $\G\bigwedge_i\psi_i$
subformula where at least one $\psi_i$ is temporal. We call such
formulae \emph{mergeable}. We first evaluate how likely we can obtain
a formula that is mergeable in Section~\ref{sec:mergeability} and then
we present the impact of our merging technique on mergeable formulae
in Section~\ref{sec:comparison}. We consider formulae that come from
two sources.

%\begin{description}
%\item[literature] 
(i) We use formulae collected from
literature~\cite{dwyer.98.fmsp, pelanek.07.spin, etessami.00.concur,
somenzi.00.cav, holecek.04.tr} that can be obtained using the tool
\texttt{genltl} from SPOT~\cite{duret.13.atva}. For each such a
formula, we added its negation, simplified all the formulae and
removed duplicates and formulae equivalent to $\true$ or $\false$. The
resulting benchmark set contains 221 formulae.

%\item[random]
(ii) We use the tool \texttt{randltl} from SPOT to generate random
formulae. We generate formulae with up to five atomic propositions and
with tree size equal to 15 (the default settings of \texttt{randltl})
before simplifications. We consider 4 different sets of random
formulae. The generator allows the user to specify \emph{priority} for
each LTL operator which determines how likely the given operator
appears in the generated formulae. By default, all operators (boolean
and temporal) have priority 1 in \texttt{randltl}. For the sets
\emph{rand1}, \emph{rand2}, and \emph{rand4}, the number indicates the
priority that was used for $\F$ and $\G$. The last set called
\emph{randfg} uses priority 2 for $\F$ and $\G$ and sets 0 to all
other temporal operators. For Section~\ref{sec:mergeability} we
generated 1000 formulae for each priority setting, for
Section~\ref{sec:comparison} we generate for each priority setting
1000 formulae that are mergeable (and throw away the rest).
%\end{description}

\begin{table}[tb]
\caption{Reference table of tools used for the experimental evaluation.}
\label{tab:tools}
\centering
\begin{tabular}{lr@{~~~}l}
\toprule
tool        & version&homepage\\
\midrule
LTL3BA      & 1.1.3  & \url{https://sourceforge.net/projects/ltl3ba}\\
LTL3TELA    & 2.1.0  & \url{https://github.com/jurajmajor/ltl3tela}\\
SPOT library& 2.7.4  & \url{https://spot.lrde.epita.fr}\\
\bottomrule
\end{tabular}
\end{table}

%\begin{table}[b!]
%\caption{Reference table of commands used to generate the automata.
%The string \texttt{\%f} is a placeholder for the input formula.}
%\label{tab:commands}
%\centering
%\begin{tabular}{l@{~~~}l}
%\toprule
%tool        & command\\
%\midrule
%LTL3BA          & \texttt{ltl3ba -H1 -f \%f} \\
%basic           & \texttt{ltl3tela -p1 -F0 -G0 -i1 -X1 -f \%f} \\
%$\F$-merging    & \texttt{ltl3tela -p1 -F0 -i1 -X1 -f \%f} \\
%$\FG$-merging   & \texttt{ltl3tela -p1 -i1 -X1 -f \%f} \\
%\bottomrule
%\end{tabular}
%\end{table}

%The tools LTL3BA and LTL3TELA were executed on all the benchmarks and
%the results were collected with the help of \texttt{ltlcross},
% another
%tool of SPOT.

\subsection{Mergeability}\label{sec:mergeability}

The set of formulae from literature contains mainly quite simple
formulae. As a result, only 24 out of the 221 formulae are mergeable.
For \emph{rand1, rand2, rand4,} and \emph{randfg} we have 302, 488,
697, and 802 mergeable formulae out of 1000, respectively.
Consistently with intuition, the ratio of mergeable formulae increases
considerably with $\F$ and $\G$ being more frequent.

\subsection{Comparison on Mergeable formulae}\label{sec:comparison}

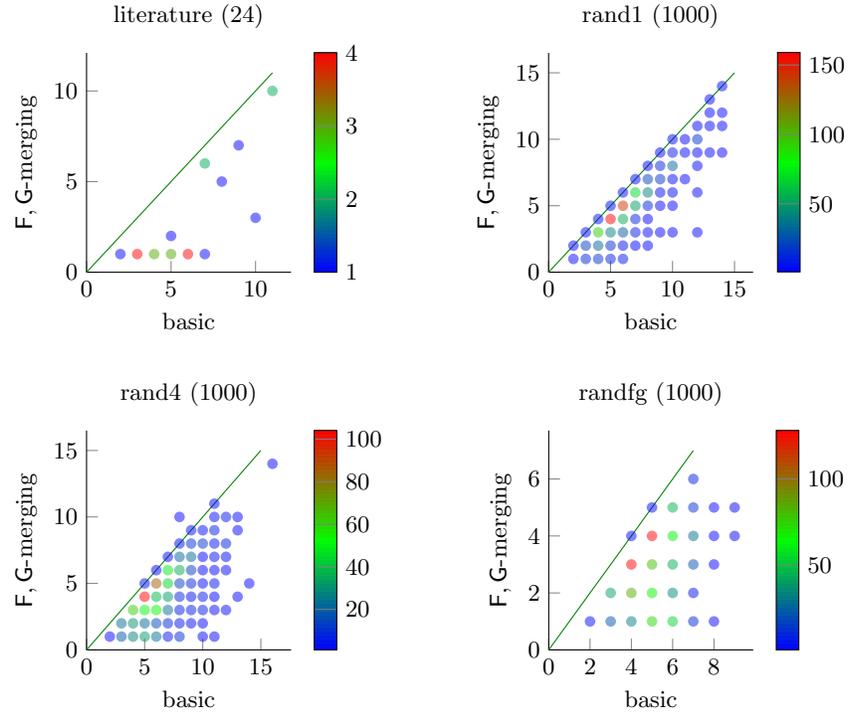
\begin{figure}[t]
\begin{center}
  \begin{tikzpicture}
  \node[matrix, column sep=30pt, row sep=1.5em](m) {
  %\begin{tikzpicture}
\pgfplotsset{every axis legend/.append style={
cells={anchor=west},
draw=none,
}}
\pgfplotsset{colorbar/width=.3cm}
\pgfplotsset{title style={align=center,
                        font=\small}}
\pgfplotsset{compat=1.14}
\begin{axis}[
xmin=0,ymin=0,
colorbar,
colormap={example}{
  color(0)=(blue)
  color(500)=(green)
  color(1000)=(red)
},
%thick,
axis x line* = bottom,
axis y line* = left,
width=4.3cm, height=4.5cm, 
xlabel={basic},
ylabel={$\FG$-merging},
cycle list={%
{darkgreen, solid},
{blue, densely dashed},
{red, dashdotdotted},
{brown, densely dotted},
{black, loosely dashdotted}
},
title={literature (24)},
]
\addplot[
    scatter, scatter src=explicit, 
    only marks, fill opacity=0.5,
    draw opacity=0] coordinates
    {(2,1) [1]
 (3,1) [4]
 (4,1) [3]
 (5,1) [3]
 (5,2) [1]
 (6,1) [4]
 (7,1) [1]
 (7,6) [2]
 (8,5) [1]
 (9,7) [1]
 (10,3) [1]
 (11,10) [2]
};%
\addplot[darkgreen,domain=0:11]{x};%
\end{axis}
%\end{tikzpicture}
&
  %\begin{tikzpicture}
\pgfplotsset{every axis legend/.append style={
cells={anchor=west},
draw=none,
}}
\pgfplotsset{colorbar/width=.3cm}
\pgfplotsset{title style={align=center,
                        font=\small}}
\pgfplotsset{compat=1.14}
\begin{axis}[
xmin=0,ymin=0,
colorbar,
colormap={example}{
  color(0)=(blue)
  color(500)=(green)
  color(1000)=(red)
},
%thick,
axis x line* = bottom,
axis y line* = left,
width=4.3cm, height=4.5cm, 
xlabel={basic},
ylabel={$\FG$-merging},
cycle list={%
{darkgreen, solid},
{blue, densely dashed},
{red, dashdotdotted},
{brown, densely dotted},
{black, loosely dashdotted}
},
title={rand1 (1000)},
]
\addplot[
    scatter, scatter src=explicit, 
    only marks, fill opacity=0.5,
    draw opacity=0] coordinates
    {(2,1) [3]
 (2,2) [1]
 (3,1) [12]
 (3,2) [21]
 (3,3) [2]
 (4,1) [15]
 (4,2) [32]
 (4,3) [95]
 (4,4) [4]
 (5,1) [10]
 (5,2) [14]
 (5,3) [39]
 (5,4) [159]
 (5,5) [4]
 (6,1) [7]
 (6,2) [5]
 (6,3) [29]
 (6,4) [54]
 (6,5) [142]
 (6,6) [4]
 (7,2) [1]
 (7,3) [8]
 (7,4) [11]
 (7,5) [48]
 (7,6) [68]
 (7,7) [2]
 (8,2) [1]
 (8,3) [3]
 (8,4) [1]
 (8,5) [12]
 (8,6) [40]
 (8,7) [21]
 (8,8) [1]
 (9,3) [1]
 (9,5) [2]
 (9,6) [5]
 (9,7) [17]
 (9,8) [12]
 (9,9) [2]
 (10,3) [1]
 (10,5) [1]
 (10,6) [3]
 (10,7) [3]
 (10,8) [33]
 (10,9) [9]
 (10,10) [3]
 (11,7) [1]
 (11,9) [3]
 (11,10) [1]
 (12,3) [1]
 (12,6) [1]
 (12,8) [2]
 (12,9) [1]
 (12,10) [15]
 (12,11) [1]
 (13,9) [3]
 (13,11) [2]
 (13,12) [1]
 (13,13) [1]
 (14,9) [1]
 (14,11) [1]
 (14,12) [2]
 (14,14) [2]
};%
\addplot[darkgreen,domain=0:15]{x};%
\end{axis}
%\end{tikzpicture}
\\
  %\begin{tikzpicture}
\pgfplotsset{every axis legend/.append style={
cells={anchor=west},
draw=none,
}}
\pgfplotsset{colorbar/width=.3cm}
\pgfplotsset{title style={align=center,
                        font=\small}}
\pgfplotsset{compat=1.14}
\begin{axis}[
xmin=0,ymin=0,
colorbar,
colormap={example}{
  color(0)=(blue)
  color(500)=(green)
  color(1000)=(red)
},
%thick,
axis x line* = bottom,
axis y line* = left,
width=4.3cm, height=4.5cm, 
xlabel={basic},
ylabel={$\FG$-merging},
cycle list={%
{darkgreen, solid},
{blue, densely dashed},
{red, dashdotdotted},
{brown, densely dotted},
{black, loosely dashdotted}
},
title={rand4 (1000)},
]
\addplot[
    scatter, scatter src=explicit, 
    only marks, fill opacity=0.5,
    draw opacity=0] coordinates
    {(2,1) [1]
 (3,1) [18]
 (3,2) [23]
 (4,1) [27]
 (4,2) [30]
 (4,3) [63]
 (5,1) [32]
 (5,2) [22]
 (5,3) [51]
 (5,4) [104]
 (5,5) [3]
 (6,1) [23]
 (6,2) [15]
 (6,3) [54]
 (6,4) [38]
 (6,5) [85]
 (6,6) [1]
 (7,1) [5]
 (7,2) [8]
 (7,3) [23]
 (7,4) [32]
 (7,5) [44]
 (7,6) [48]
 (7,7) [1]
 (8,1) [9]
 (8,2) [4]
 (8,3) [6]
 (8,4) [8]
 (8,5) [13]
 (8,6) [36]
 (8,7) [21]
 (8,8) [2]
 (8,10) [1]
 (9,2) [3]
 (9,3) [6]
 (9,4) [3]
 (9,5) [8]
 (9,6) [13]
 (9,7) [22]
 (9,8) [10]
 (9,9) [3]
 (10,1) [1]
 (10,2) [1]
 (10,3) [6]
 (10,4) [5]
 (10,5) [2]
 (10,6) [1]
 (10,7) [1]
 (10,8) [11]
 (10,9) [3]
 (11,1) [1]
 (11,3) [2]
 (11,4) [1]
 (11,5) [4]
 (11,6) [1]
 (11,7) [3]
 (11,8) [4]
 (11,9) [6]
 (11,10) [3]
 (11,11) [1]
 (12,3) [3]
 (12,6) [2]
 (12,7) [2]
 (12,8) [7]
 (12,10) [5]
 (13,4) [1]
 (13,9) [1]
 (13,10) [1]
 (14,5) [1]
 (16,14) [1]
};%
\addplot[darkgreen,domain=0:15]{x};%
\end{axis}
%\end{tikzpicture}
&
  %\begin{tikzpicture}
\pgfplotsset{every axis legend/.append style={
cells={anchor=west},
draw=none,
}}
\pgfplotsset{colorbar/width=.3cm}
\pgfplotsset{title style={align=center,
                        font=\small}}
\pgfplotsset{compat=1.14}
\begin{axis}[
xmin=0,ymin=0,
colorbar,
colormap={example}{
  color(0)=(blue)
  color(500)=(green)
  color(1000)=(red)
},
%thick,
axis x line* = bottom,
axis y line* = left,
width=4.3cm, height=4.5cm, 
xlabel={basic},
ylabel={$\FG$-merging},
cycle list={%
{darkgreen, solid},
{blue, densely dashed},
{red, dashdotdotted},
{brown, densely dotted},
{black, loosely dashdotted}
},
title={randfg (1000)},
]
\addplot[
    scatter, scatter src=explicit, 
    only marks, fill opacity=0.5,
    draw opacity=0] coordinates
    {(2,1) [3]
 (3,1) [22]
 (3,2) [41]
 (4,1) [32]
 (4,2) [85]
 (4,3) [128]
 (4,4) [3]
 (5,1) [74]
 (5,2) [68]
 (5,3) [81]
 (5,4) [128]
 (5,5) [2]
 (6,1) [54]
 (6,2) [35]
 (6,3) [44]
 (6,4) [62]
 (6,5) [45]
 (7,1) [11]
 (7,2) [5]
 (7,3) [21]
 (7,4) [29]
 (7,5) [13]
 (7,6) [8]
 (8,1) [1]
 (8,3) [1]
 (8,4) [1]
 (8,5) [1]
 (9,4) [1]
 (9,5) [1]
};%
\addplot[darkgreen,domain=0:7]{x};%
\end{axis}
%\end{tikzpicture}
\\
  };
  \end{tikzpicture}
\end{center}
\vspace{-1em}
\caption{Effect of $\FG$-merging on SLAA size for mergeable formulae. A dot represents the number of states of
the SLAA produced by $\FG$-merging ($y$-axis) and by the basic
translation ($x$-axis) for the same formula. The color of the dot
reflects the number of dots at the position.}
\label{fig:m-impact}
\end{figure}

Table~\ref{tab:m-impact} shows the cumulative numbers of states and
acceptance marks of SLAA produced by LTL3BA and the translations
presented in Sections~\ref{sec:basic},~\ref{sec:f-merging},
and~\ref{sec:fg-merging} as implemented in LTL3TELA for mergeable
formulae. The scatter plots in Figure~\ref{fig:m-impact} offer more
details on the improvement (in the number of states) of $\FG$-merging
over basic translation. The missing plot for rand2 was omitted for
space reasons as it is very similar to the one for rand1.

Merging of states has a positive impact also on the type of the
produced automaton as it removes both some universal and
nondeterministic branching from the automaton.
Table~\ref{tab:alt-det-nondet} shows for each set of formulae and each
translation the number of automata that have no universal branching (right part) and that are even deterministic (left part).

\begin{table}[t]
\caption{Comparison of LTL to SLAA translations on mergeable formulae.}
\label{tab:m-impact}
\centering
\setlength{\tabcolsep}{3pt}
\begin{tabular}{lrrrrrcrrrrr}
\toprule
{} & \multicolumn{5}{c}{states} & ~ & \multicolumn{5}{c}{acceptance marks} \\
\cmidrule(lr){2-6} \cmidrule(lr){8-12}
{} & lit & rand1 & rand2 & rand4 & randfg & ~ & lit & rand1 & rand2 & rand4 & randfg \\
\# of form. & 24 & 1000 & 1000 & 1000 & 1000 & & 24 & 1000 & 1000 & 1000 & 1000 \\
\midrule
LTL3BA        &        140 &  6253 &  6313 &  6412 &   5051 & &       24 &  1000 &  1000 &  1000 &   1000 \\
basic         &        140 &  6234 &  6287 &  6393 &   5051 & &       24 &   997 &  1000 &  1000 &   1000 \\
$\F$-merging  &        110 &  5418 &  5296 &  5231 &   3926 & &       46 &  1160 &  1244 &  1347 &   1343 \\
$\FG$-merging &         65 &  4595 &  4300 &  4015 &   2744 & &       98 &  2971 &  3317 &  3677 &   2978 \\
\bottomrule
\end{tabular}
\end{table}

One can observe in the tables that the basic translation produces
similar SLAA as LTL3BA. Furter, the $\F$-merging and $\FG$-merging
translations bring gradual improvement both in the number of states
and in the numbers of deterministic and nonalternating automata in
comparison to the basic translation on all sets of formulae. The ratio
of states that can be saved by merging grows with the increasing
occurrence of $\F$ and $\G$ operators up to 45\% (randfg) in the
benchmarks that use randomly generated formulae.

The scatter plots reveal that most cases fit into the category
\emph{the $\FG$-merging translation saves up to 3 states}. But there
are also cases where the $\FG$-merging translation reduces the
resulting SLAA to 1 state only while the basic translation needs 8 or
even more states. However, we still have one case where the basic
translation produces smaller automaton than the $\FG$-merging
(see Figure~\ref{fig:m-impact}, rand4) and $\F$-merging translations.

Table~\ref{tab:alt-det-nondet} confirms by numbers that the $\F$- and
$\FG$-merging translations often build automata with fewer branching.
The numbers of deterministic and nonalternating (without universal
branching) automata are especially appealing for the $\FG$-merging
translation on formulae from the set \emph{randfg}.

\begin{table}[t]
\caption{Number of alternating automata that use only existential
branching. The numbers of formulae are the same as in
Table~\ref{tab:m-impact} for each set.} \label{tab:alt-det-nondet}
\centering \setlength{\tabcolsep}{3pt}
\begin{tabular}{lrrrrrcrrrrr}
\toprule
{} & \multicolumn{5}{c}{deterministic} & ~ & \multicolumn{5}{c}{nonalternating} \\
\cmidrule(lr){2-6} \cmidrule(lr){8-12}
{} & lit & rand1 & rand2 & rand4 & randfg & ~ & lit & rand1 & rand2 & rand4 & randfg \\
\midrule
LTL3BA        &          0 &     5 &     2 &     0 &      0 & &        6 &   148 &   114 &    89 &    144 \\
basic         &          0 &     5 &     2 &     0 &      0 & &        6 &   148 &   114 &    89 &    144 \\
$\F$-merging  &          2 &    64 &    53 &    40 &     73 & &        6 &   171 &   146 &   119 &    192 \\
$\FG$-merging &         10 &   133 &   126 &   124 &    217 & &       18 &   356 &   367 &   385 &    603 \\
\bottomrule
\end{tabular}
\end{table}

On the downside, the presented translations produce SLAA with more
acceptance marks than the translation of LTL3BA. This is the price we
pay for small automata. Basic translation sometimes uses 0 acceptance
marks if there is no $\F$ or $\U$ operator.

%\begin{figure}[t!]
%\centering
%\begin{tikzpicture}
%\node[matrix, column sep=30pt](m) {
%%\input{sc_rand1.tex}&
%\input{sc_rand2.tex}&
%\input{sc_rand4.tex}&
%\input{sc_randfg.tex}\\
%};
%\end{tikzpicture}
%\vspace{-1em}
%\caption{Effect of $\FG$-merging on SLAA size for rand4 and randfg formulae.} \label{fig:m-impact-rand}
%\end{figure}

%}}}
%{{{ conclusion

\section{Conclusion}

We have presented a novel translation of LTL to self-loop alternating
automata (SLAA) with Emerson-Lei acceptance condition.  To our best
knowledge, it is the first translation of LTL producing SLAA with
other than Büchi or co-Büchi acceptance. Our experimental results
demonstrated that the expressive acceptance condition allows to
produce substantially smaller SLAA comparing to these
produced by LTL3BA when $\F$ or $\G$ operators appear in the translated formula. %

This work opens doors for research of algorithms processing SLAA with
Emerson-Lei acceptance, in particular the algorithms transforming
these SLAA to nonalternating automata with various degrees of
determinism: nondeterministic, deterministic, or semi-deterministic
(also known as limit-deterministic) automata. Our implementation can
serve as a natural source of such automata and already helped to build
a tool that can produce small deterministic and nondeterministic TELA.

%}}}

\clearpage
\sloppy
\printbibliography

\iffullversion
\newpage
\appendix

%The following appendix is not in the official version of this paper published at ICTAC 2019.

%{{{ Correctness of the F,G-merging translation

\section{Correctness and Complexity of the $\FG$-Merging Translation}
\label{app:proof}

{
\renewcommand{\thetheorem}{\ref{thm:fg-merging}}

\begin{theorem}
  Let $\varphi$ be an LTL formula and let $\mA_\varphi$ be the
  corresponding SLAA built by the $\FG$-merging translation. Then
  $L(\mA_\varphi) = L(\varphi)$.  Moreover, the number of states of
  $\mA_{\varphi}$ is linear to the size of $\varphi$ and the number of
  acceptance marks is at most exponential to the size of $\varphi$.
\end{theorem}
\addtocounter{theorem}{-1}
} % note: these braces are here to take advantage of LaTeX scoping
  % \thetheorem is returned to its rightful definition outside of this group

\begin{proof}
We prove the correctness part of the theorem by structural induction
in both directions. We suppose that, for all strict subformulae $\psi$
of $\varphi$, the SLAA for $\psi$, which we denote $\mA_\psi$, is
correct (accepts exactly $\mathcal{L}(\psi)$) and also corresponding
SLAA for all subformulae of $\psi$ are correct in the same sense.

In the proof, the following notation is used. Let $T$ be a
multitransition and $S$ be a set of states. Then $T[S] = \{t =
(s,\alpha,M,C)\mid t\in T\text{ and }s\in S\}$ is the set of
transitions from $T$ with a source from $S$, and $T[\bar{S}] = T
\smallsetminus T[S]$. For $K \in \overline{\psi}$, we again use the
notation $\psi_K = \bigwedge_{\psi' \in K} \psi'$.

To prove $\mathcal{L}(\varphi) \subseteq \mathcal{L}(\mA_\varphi)$,
let $w$ be a word such that $w \models \varphi$. We find an accepting
run $\rho = T_0T_1\ldots$ of $\mA_\varphi$ over $w$ for each possible
form of $\varphi$.
We will often compose runs of automata for subformulae on suffixes of
$w$. Let $0 \leq i \leq j$ be two indices. We say that two runs
$T_0^iT_1^i\ldots$ and $T_0^jT_1^j\ldots$ on $w\suf{i}$ and
$w\suf{j}$, respectively, are \emph{synchronized} iff for each $k \geq
j$ we have that $T_k = T^i_{k-i} \cup T^j_{k-j}$ is a multitransition
(which means, there is at most one $(s,w_k,M,C)\in T_k$ for each state
$s$).

\begin{proposition}\label{prop:sync}
Let $w\suf{i}$ and $w\suf{j}$ be two suffixes of some
$w\in\Sigma^\omega$ and $\mA$ an SLAA such that both $w\suf{i}$ and
$w\suf{j}$ are in $\lang{\mA}$. Then there always exist accepting runs
over $w\suf{i}$ and $w\suf{j}$ that are synchronized. This extends
naturally also to runs of different automata built by our inductive
definition.
\end{proposition}

In the following, when we choose multiple accepting runs over suffixes
of some word, we always assume they are synchronized. This is a valid
assumption due to Proposition~\ref{prop:sync}.

If $\varphi$ is one of $\true$, $a$ or $\lnot a$ where $a \in AP$, we
have $T_0 = \left\{\left(\varphi, w_0, \emptyset,
\emptyset\right)\right\}$ and $T_i = \emptyset$ for $i \geq 1$. The
transition in $T_0$ exists in $\mA_\varphi$ by definition and the run
is accepting since it has no infinite branch. If $\varphi \equiv
\false$, then no word satisfies $\varphi$ and there is nothing to
prove.

Let $\varphi \equiv \psi_1 \lor \psi_2$ \ldots, we can assume that $w
\models \psi_1$ (the proof is analogous in the other case). Therefore,
there is an accepting run $\sigma$ of $\mA_{\psi_1}$ over $w$, and we
define $\rho$ as $\rho=\{(\varphi,w_0,\emptyset,C)\}T_1T_2\ldots$
where $\sigma = \{(\psi_1,w_0,M,C)\}T_1T_2$. The run $\rho$ is
accepting as it has branches isomorphic to branches of $\sigma$.

Let $\varphi \equiv \psi_1 \land \psi_2$. Then we can assume two
synchronized runs $R_0R_1\ldots$ of $\mA_{\psi_1}$ and $S_0S_1\ldots$
of $\mA_{\psi_2}$ over $w$ with $R_0=\{(\psi_1,w_0,M_1,C_1)\}$ and
$S_0=\{(\psi_2,w_0,M_2,C_2)\}$. We define $T_0 =
\{(\varphi,w_0,\emptyset,C_1 \cup C_2)\}$ and $T_i = R_i \cup S_i$ for
$i > 0$. The run $\rho$ is a union of branches of the two runs and
thus is accepting.

If $\varphi \equiv \X\psi$, then we create $\rho$ as $\{(\varphi,
w_0, \emptyset, \{\psi\})\}T_1T_2\ldots$ where $T_1T_2\ldots$ is an
accepting run of $\mA_\psi$ over $w\suf{1}$.

Let $\varphi \equiv \psi_1 \U \psi_2$ where $\psi_1 \not\equiv \true$
and let $k \geq 0$ be the smallest index such that $w_{k..} \models
\psi_2$. We know that $w_{j..} \models \psi_1$ for all $j < k$ as
$w\models \varphi$, and thus for each such $j$ there exist an
accepting run $\sigma^j = S^j_0S^j_1\ldots$ of $\mA_{\psi_1}$ over
$w_{j..}$. Further, there is an accepting run $\sigma^k = S^k_0S^k_1
\ldots$ of $\mA_{\psi_2}$ over $w_{k..}$. We define the
multitransitions of $\rho$ as follows.
\[
T_i = \bigcup_{\substack{0 \leq j < i\\j < k}}\!\! S^j_{i-j}
\cup \begin{cases}
  \left\{
    \left(\varphi, w_i, \left\{\gm_\varphi\right\},
      C \cup \left\{\varphi\right\}
    \right) \mid
    \left(\psi_1, w_i, M, C\right) \in S^i_0
  \right\}
  & i < k \\
  \left\{
    \left(\varphi, w_i, \emptyset, C\right) \mid
    \left(\psi_2, w_i, M, C\right) \in S^k_0
  \right\}
  & i = k \\
  S^k_{i-k}
  & i > k \\
\end{cases}
\]
The run $\rho$ contains the branches of all $\sigma^j$ for $j\leq k$,
some of them prefixed with finitely many $\gm_\varphi$-labelled edges
between nodes of the form $(\varphi,i)$. Since $\gm_\varphi$ only
appears in the first $k$ multitransitions, $\Fin\gm_\varphi$ is
satisfied, the rest of the acceptance formula is satisfied as we reuse
the branches of accepting runs.

If $\varphi \equiv \F\psi$, we again denote by $k$ the smallest index
such that $w_{k..} \models \psi_K$ for some $K \in \overline{\psi}$.
Let $M_K = \{ \om^{K'}_\varphi \mid K \neq K' \in \overline{\psi}\}$.
By the induction hypothesis, there exists a run $\sigma =
S_0S_1\ldots$ of $\mA_{\psi_K}$ over $w_{k..}$. Let $l$ be the minimal
index $l \geq k$ such that $K \not \subseteq \range(S_l[K])$ if exists
and $l = \infty$ otherwise. In the following definition, we
exceptionally redefine $S_0[K] = S_0$ and $S_0[\bar{K}] = \emptyset$
in the special case of $k = 0$. To mitigate the implementation of the
trick with looping transitions of $\psi_K$, for each $k \leq i < l$ we
define $C_i = \range(S_i[K])\smallsetminus K$. We can finally define
the multitransitions of $\rho$ as follows.

\[
T_i =
\begin{cases}
  \left\{ \left(
    \varphi, w_i, \left\{\gm_\varphi\right\}, \{\varphi\}
  \right) \right\}
  & i < k\\
  \{(\varphi,w_i,M_K,C_i \cup \{\varphi\})\} \cup S_i[\bar{K}]
  & k \leq i < l\\
  \{(\varphi,w_i,\emptyset,\range(S_i[K]))\}\cup S_i[\bar{K}]
  & i = l\\
  S_i
  & i > l\\
\end{cases}
\]

The run $\rho$ is accepting because $\gm_\varphi$ only appears in
first $k$ multitransitions, $\om^K_\varphi$ does not appear anywhere
in the run, and the rest of the condition inherited from automata for
subformulae is trivially satisfied by the branch that stays in
$\varphi$ forever and is satisfied on the other branches as they only contain marks of $\mA_{\psi_K}$ and they mimic
(using $S_i[\bar{K}]$) the accepting run $\sigma$.

Let $\varphi \equiv \psi_1 \R \psi_2$ such that $\psi_1 \not\equiv
\false$ or $\psi_2$ is not a conjunction of temporal and state
formulae. Let $k$ be the smallest number such that $w_{k..} \models
\psi_1 \land \psi_2$ and that for all $i < k$ we have $w_{i..} \models
\psi_2$. If no such $k$ exists, we set $k = \infty$. Similarly as for
the $\U$-case, from the induction principle we have accepting runs
$\sigma^i = S^i_0S^i_1\ldots$ of $\mA_{\psi_2}$ over $w_{i..}$ for $0
\leq i \leq k$ and an accepting run $\sigma = S_0S_1 \ldots$ of
$\mA_{\psi_1}$ over $w_{k..}$. The multitransitions of $\rho$ are
defined as follows.
\[
T_i = \bigcup_{j=0}^{i-1} S^j_{i-j} \cup
\begin{cases}
  \{(\varphi, w_i, \emptyset, C \cup \{\varphi\} \mid
  (\psi_2, w_i, M, C) \in S^i_0 ) \}
  & i < k \\
  \{(\varphi, w_k, \emptyset, \range(S_0)\cup\range(S^k_0) )\}
  & i = k \\
  S_{i-k}
  & i > k \\
\end{cases}
\]

Finally, let $\varphi \equiv \G\bigwedge_{\psi \in K} \psi$ where each
$\psi \in K$ is a temporal or state formula. For every $\psi \in K$
and every $i \geq 0$ we know that $w\suf{i}\models\psi$ and thus by
induction hypothesis we have accepting runs $\sigma^{\psi,i} =
S^{\psi,i}_{0}S^{\psi,i}_{1}\ldots$ of $\mA_\psi$ over $w_{i..}$.
Further, we may assume that these runs are synchronized due to
Proposition~\ref{prop:sync}, and thus $S^{\psi,i}_0 \subseteq
S^{\psi,i-j}_j$ for all $j \leq i$. Similarly to the $\F\psi'$ case,
we define $C^\psi_i = \range(S^{\psi,i}_0)\smallsetminus \{\psi\}$.
Moreover, assuming $S^{\psi,i}_0 = \{(\psi,w_i,M,C)\}$, we define
$M^\psi_i = \{\bsq_\psi \}$ if $S^{\psi,i}_0$ is $\psi$-escaping and
$\psi \in \mF_\varphi \cup \mU_\varphi$, and $M^\psi_i = M$ otherwise.

\[
\begin{aligned}
T_i = \left\{ \left(
  \varphi, w_i, M_i, C_i
\right) \right\} \cup 
  \bigcup_{\substack{\psi\in K\\0\leq j < i}}
  S^{\psi,j}_{i-j}[\overline{\{\psi\}}]\text{, where}
  \\
M_i = \bigcup_{\psi\in K} M^\psi_i \hspace{1cm}
C_i = \{\varphi\} \cup \bigcup_{\psi\in K} C^\psi_i
\end{aligned}
\]

The acceptance formula is satisfied by $\rho$ as, for each $\psi\in K
\cap (\mU \cup \mF)$ the marks $\gm_\psi$ (and some $\om_\psi^S$)
appear only in finitely many multitransitions of $\rho$, or there are
infinitely many $\psi$-escaping multitransitions and we have
$\Inf\bsq_\psi$ satisfied by the branch staying in $\varphi$. The rest
of the formula is satisfied by $\rho$ because it was satisfied by all
the branches of runs $\sigma^\psi_i$.

\medskip

To prove $\mathcal{L}(\mA_\varphi) \subseteq \mathcal{L}(\varphi)$, we
show that if there exists an accepting run $\rho = T_0T_1\ldots$ of
$\mA_{\varphi}$ over $w$, then $w \models \varphi$. Again we prove
this using structural induction with the induction hypothesis assuming
that the above holds for every strict subformula of $\varphi$. For a
run $\sigma = T_0T_1\ldots$ of $\mA_\varphi$ and a set of states $C$
and an index $i\geq 0$ we use $T_i[C]\ldots$ to describe the run
$\sigma' = T_i[C]T_{i+1}[\range(T_i[C])]\ldots$ of $\mA_\psi$ over
$w\suf{i}$ where $\psi = \bigwedge_{\psi'\in C}\psi'$. If $\sigma$ is
accepting, $\sigma'$ is also accepting as it contains only suffixes of
branches of $\sigma$.

The statement is trivially true for $\varphi \equiv \true$ (every word
satisfies $\true$) and $\varphi \equiv \false$ (there cannot exist an
accepting run over $\mA_\false$).

If $\varphi \equiv a$, then surely $T_0$ is an $w_0$-labelled
multitransition such that $a \in w_0$, hence $w \models \varphi$.
Similar argument applies to $\varphi \equiv \lnot a$.

Let $\varphi \equiv \psi_1 \lor \psi_2$. Then $\rho$ is (up to $\dom(T_0)$)
equivalent to (also accepting) run of $\mA_{\psi_1}$ or $\mA_{\psi_2}$ over $w$.
Therefore $w \models \psi_1$ or $w \models \psi_2$, so $w \models \varphi$.

If $\varphi \equiv \psi_1 \land \psi_2$, then, by definition of $\Delta$,
$\rho$ consists exactly of branches of runs of $\mA_{\psi_1}$ and $\mA_{\psi_2}$,
again with the only difference in $\dom(T_0)$, all of which have to be
accepting. Therefore, by the induction hypothesis, $w \models \psi_1$ and $w
\models \psi_2$, so $w \models \varphi$.

Let $\varphi \equiv \X\psi$. By the definition of $\Delta$, $T_1T_2\ldots$
is an accepting run of $\mA_\psi$ over $w_{1..}$, thus $w_{1..} \models \psi$
and so $w \models \varphi$.

Let $\varphi \equiv \psi_1 \U \psi_2$. Since $\rho$ is accepting but,
by the definition of $\Delta$, no $\bsq_\varphi$ or $\om^K_\varphi$
appear in the run. Therefore $\Fin\gm_\varphi$ has to hold. We can see
that every outgoing transition of $\varphi$ is looping if and only if
its acceptance label contains $\gm_\varphi$, so there exists some $k
\in \mathbb{N}$ such that, after $k$ steps, the run escapes from
$\varphi$ and, by the definition of $\Delta$, the run
$T_kT_{k+1}\ldots$ is an accepting run of $\mA_{\psi_2}$ over
$w_{k..}$ and thus $w\suf{k}\models\psi_2$ by induction hypothesis.
Moreover, for each $i < k$ the multitransition $T_i$ contains (by
definition of $\Delta$) a transition
$(\varphi,w_i,\{\gm_\varphi\},\{\varphi\}\cup C)$ such that
$\{(\psi_1, w_i, M, C)\}T_{i+1}[C]\ldots$ for some $M$ is an accepting
run of $\mA_{\psi_1}$ over $w\suf{i}$, so $w\suf{i} \models \psi_1$.

If $\varphi \equiv \F\psi$, then again no $\bsq_\varphi$ appears in the run,
so both $\Fin\gm_\varphi$ and $\Fin\om^K_\varphi$ for some $K \in
\overline{\psi}$ holds. Therefore there exists $k \in \mathbb{N}$ such that
neither $\gm_\varphi$ nor $\om^K_\varphi$ appears in $T_kT_{k+1}\ldots$.
This happens because of two reasons: either the run eventually leaves the
state $\varphi$ using some transition from $\psi_{K'}$ (where $K' \in
\overline{\psi}$), or we only use transitions from $\psi_K$ (with $\psi_K$
replaced by $\varphi$). In both cases, $w_k \models \psi$ so $w \models
\varphi$.

If $\varphi \equiv \psi_1 \R \psi_2$, each $T_i$ that contains
$\varphi$ contains a transition $(\varphi,w_i,\emptyset,C_1\cup
C_2)$ such that $\{(\psi_2, w_i, M, C_2)\}T_{i+1}[C_2]\ldots$ for some
$M$ is an accepting run of $\mA_{\psi_2}$ over $w\suf{i}$, hence
$w\suf{i} \models \psi_2$. If $C_1 = \{\varphi\}$ for all $i$, we have
that $w\models\varphi$. Otherwise, there is a minimal $k\geq 0$ such
that $\{(\psi_1, w_k, M, C_1)\}T_{k+1}[C_1]\ldots$ for some $M$ is an
accepting run of $\mA_{\psi_1}$ over $w\suf{k}$, hence $w\suf{k}
\models \psi_1$ and again, $w \models \varphi$.

Finally, let $\varphi \equiv \G \bigwedge_{\psi \in K} \psi$ where
$\psi \in K$ are all temporal or state formulae. For each $\psi\in K$
and each $i$ we construct an accepting run $\sigma^\psi_i$ of
$\mA_\psi$ over $w\suf{i}$. By definition of $\Delta$ we know that in
each $T_i$ there is a transition of the form $(\varphi,w_i,M_i,C_i)$.
Let $C^\psi_i = \{\psi\}$ if $\bsq_\psi \notin M_i$, and $C_\psi' =
\emptyset$ otherwise. By definition of $\Delta$, there is some $C_i'
\subset C_i$ such that $t = (\psi,w_i,M_i',C_i'\cup C^\psi_i)$ is a
transition of $\mA_{\psi}$ for some $M_i'$ which is arbitrary if
$\bsq_{\psi}\in M_i$ and $M_i'\subseteq M_i$ otherwise. If $C^\psi_i =
\emptyset$, then $\sigma^\psi_i = \{t\}T_{i+1}[C_i']$ and
$\sigma^\psi_i = \{t\}(T_{i+1}[C_i']\ldots \sqcup \sigma^\psi_{i+1})$
where $\sqcup$ makes a union of corresponding multitransitions
(multitransitions on the same positions) of two runs, otherwise.

Each $\sigma^\psi_i$ consists of a subset of branches of $\rho$ (or
their suffixes) and some branches that loop in $\psi$ for a while. If
the branch stays forever in $\psi$ then its marks are exactly those
marks of $\rho$ relevant to $\psi$ (have the $\psi$ subscript) and
such branch must satisfy the accepting formula of $\mA_\psi$. Branches
that eventually leave $\psi$ have suffix that is already in $\rho$ and
thus must satisfy the acceptance formula too. Hence, $\sigma^\psi_i$
is an accepting run of $\mA_\psi$ over $w\suf{i}$ and we have that
$w\suf{i}\models\varphi$ for each $i$ and $\psi$, hence $w\models
\varphi$. 

\medskip The linear number of states of $\mA_\varphi$ with respect to
the size of $\varphi$ comes from the fact that states of $\mA_\varphi$
are subformulae of $\varphi$. Let $n$ be the number of states of
$\mA_\varphi$. Then we have at most $n$ marks of the form $\gm_\psi$,
at most $n$ marks of the form $\bsq_\psi$, and finally at most $2^n$
marks of the form $\om_\psi^K$ (the exponential blowup may be caused
only by conversion to DNF, if necessary). Overall, the number of marks
is at most exponential with respect to size of $\varphi$.  \qed
\end{proof}

%}}}
\fi

\end{document}